\documentclass{article}
\usepackage[utf8]{inputenc}
\usepackage{xcolor}
\usepackage{amsmath}
\usepackage{graphicx}
\usepackage{geometry}
\usepackage{bbding}
\usepackage{authblk}
\newgeometry{vmargin={20mm}, hmargin={15mm,20mm}}   

\title{A Survey of COVID-19 Contact Tracing Apps}
\author[1,2]{Nadeem Ahmed}
\author[1,2]{Regio A. Michelin}
\author[1,2]{Wanli Xue}
\author[3]{Sushmita Ruj}
\author[2]{Robert Malaney}
\author[1,2]{Salil S. Kanhere}
\author[1,2]{Aruna Seneviratne}
\author[1,2]{Wen Hu}
\author[1,4]{Helge Janicke}
\author[1,2]{Sanjay Jha}
\affil[1]{Cyber Security Cooperative Research Centre (CSCRC) - Australia}
\affil[2]{University of New South Wales (UNSW) - Sydney, Australia}
\affil[3]{CSIRO, Data61 - Sydney, Australia}
\affil[4]{Edith Cowan University, Perth, Australia}
\affil[ ]{Corresponding author: \textit{nadeem.ahmed@cybersecuritycrc.org.au}}

\begin{document}

\maketitle

\section*{Abstract}
The recent outbreak of COVID-19 has taken the world by surprise, forcing lockdowns and straining public health care systems. COVID-19 is known to be a highly infectious virus, and infected individuals do not initially exhibit symptoms, while some remain asymptomatic. Thus, a non-negligible fraction of the population can, at any given time, be a hidden source of transmissions. In response, many governments have shown great interest in smartphone contact tracing apps that help automate the difficult task of tracing all recent contacts of newly identified infected individuals. However, tracing apps have generated much discussion around their key attributes, including system architecture, data management, privacy, security, proximity estimation, and attack vulnerability. In this article, we provide the first comprehensive review of these much-discussed tracing app attributes. We also present an overview of many proposed tracing app examples, some of which have been deployed countrywide, and discuss the concerns users have reported regarding their usage. We close by outlining potential research directions for next-generation app design, which would facilitate improved tracing and security performance, as well as wide adoption by the population at large.

\section{Introduction}
\label{sec:Intro}
The year 2020 will be forever marked in history by the worldwide outbreak of the pandemic caused by Severe Acute Respiratory Syndrome Coronavirus 2 (SARS-CoV-2) a.k.a COVID-19. As the virus began to spread globally, the World Health Organisation (WHO) declared on the 30th of January 2020 that COVID-19 was a Public Health Emergency of International Concern (PHEIC) [1]. The virus outbreak has changed the lifestyle of everyone around the world forcing governments to mandate lockdowns, recommend self-isolation, stipulate work-from-home policies, instigate strict social distancing criteria, and deploy emergency health responses - the latter including substantial new infrastructure for the treatment and mass testing of the population at large. All these measures are aimed at decreasing the rate of spread of the virus and leading to a so-called  \emph{`flattening of the curve'}\footnote{Flattening of the curve aims to decrease the infection rate so that the available health resources remain compatible with the infected caseload.} until an approved vaccine/treatment is developed.

COVID-19 is more infectious compared to other known viruses (such as SARS and MERS) albeit with a lower mortality rate. Furthermore, a COVID-19  carrier
 can be contagious without experiencing any symptoms. Thus, by the time the carrier actually \emph{tests} positive they may have already spread the virus to many others who came in contact with them. This necessitates a process called `contact tracing,' to identify individuals who had close contact with the positive carrier, as these individuals may themselves now be infected.

 Contact tracing is normally accomplished through a manual interview of the infected individuals, conducted by the health authorities. The aim of the interview is to collect contacts the infected individual had with other individuals in the past 14-21 days (identified as the incubation period for COVID-19). The health officials can then use that information  to compute a risk-score for each of the contacts, based on the context (e.g., indoors/outdoors), duration, and proximity (distance between the contacts).
However, it is challenging for people to accurately recall each person that they may have met in the last three weeks. Besides, an infected individual might have infected many persons that they cannot identify, for example, contact with unknown persons standing in a supermarket checkout queue. Moreover, many subsequent interviews require a considerable workforce of  health officials trained in the art of manual contact tracing.

In this context, researchers have been focusing on technological solutions to automate the contact tracing process with the aim of quickly and reliably identifying  contacts that might be at significant infection risk. The ubiquity of smartphones and their ability to keep track of their location (e.g., via GPS and WiFi), along with their in-built Bluetooth interface allowing communication and proximity detection with nearby smartphones, makes them ideal devices for automated and reliable contact tracing. As a result, many smartphone contact tracing  apps have been proposed, with some already deployed.\footnote{MIT Technology Review has a list of 25 such apps that have been developed and are being used in many countries \cite{mit}}  Using the Bluetooth interface these tracing apps automatically collect the contact data of their users - data to be subsequently used in the future event of a user being identified as infected with COVID-19.


The introduction of contact tracing apps has led to a debate regarding their architecture, data management, efficacy, privacy, and security~\cite{V2020,bbc, duball, jennings, palmer, farrell}. Most of these apps claim to be privacy-preserving - meaning that they do not reveal any Personally Identifiable Information (PII), identity, or location information of the contacts without explicit user permission.
Indeed, privacy concerns associated with contact tracing apps are one of the factors that influence their adoption \cite{Redmiles}. Another primary concern of privacy advocates is the extent to which the apps can be re-purposed to track their users, and how the collected data may be used when the current pandemic ends.

This article describes the salient features of current contact tracing apps and provides insights into their privacy and security implications. Our article is based on public information about current tracing apps,  and our understanding of the evolving protocols being designed for upcoming apps.  Existing surveys on contact tracing apps~\cite{LG20, 672} are either not comprehensive  or focus only on user privacy~\cite{T20}. The work in \cite{KBS20} describes formal notions of tracing app design. Our work differs from \cite{KBS20} in that we discuss not only vulnerabilities and privacy issues in design but also implementation and usability issues. 


The remainder of this article is organised as follows. In Section \ref{sec:Arch}, we classify the tracing apps in three main architectures as centralised, decentralised, or hybrid. Section \ref{sec:Privacy} discusses the security, privacy, and data management implications of these different architectures. In Section~\ref{sec:proximity}, we highlight the Bluetooth Low Energy (BLE)-based proximity estimation technique used in most of the tracing apps. The most common attacks and vulnerabilities that could affect  tracing apps in discussed in Section~\ref{sec:Attacks}. In Section~\ref{sec:apps} a more detailed analysis of the most widely discussed apps is presented, and some common user concerns associated with these apps analysed in Section~\ref{sec:commonconcerns}.  In Section~\ref{sec:future} potential future research directions are given. Finally, Section \ref{sec:concl} concludes this article.

\begin{figure}[htb]
    \centering
    \includegraphics[width=0.75\textwidth]{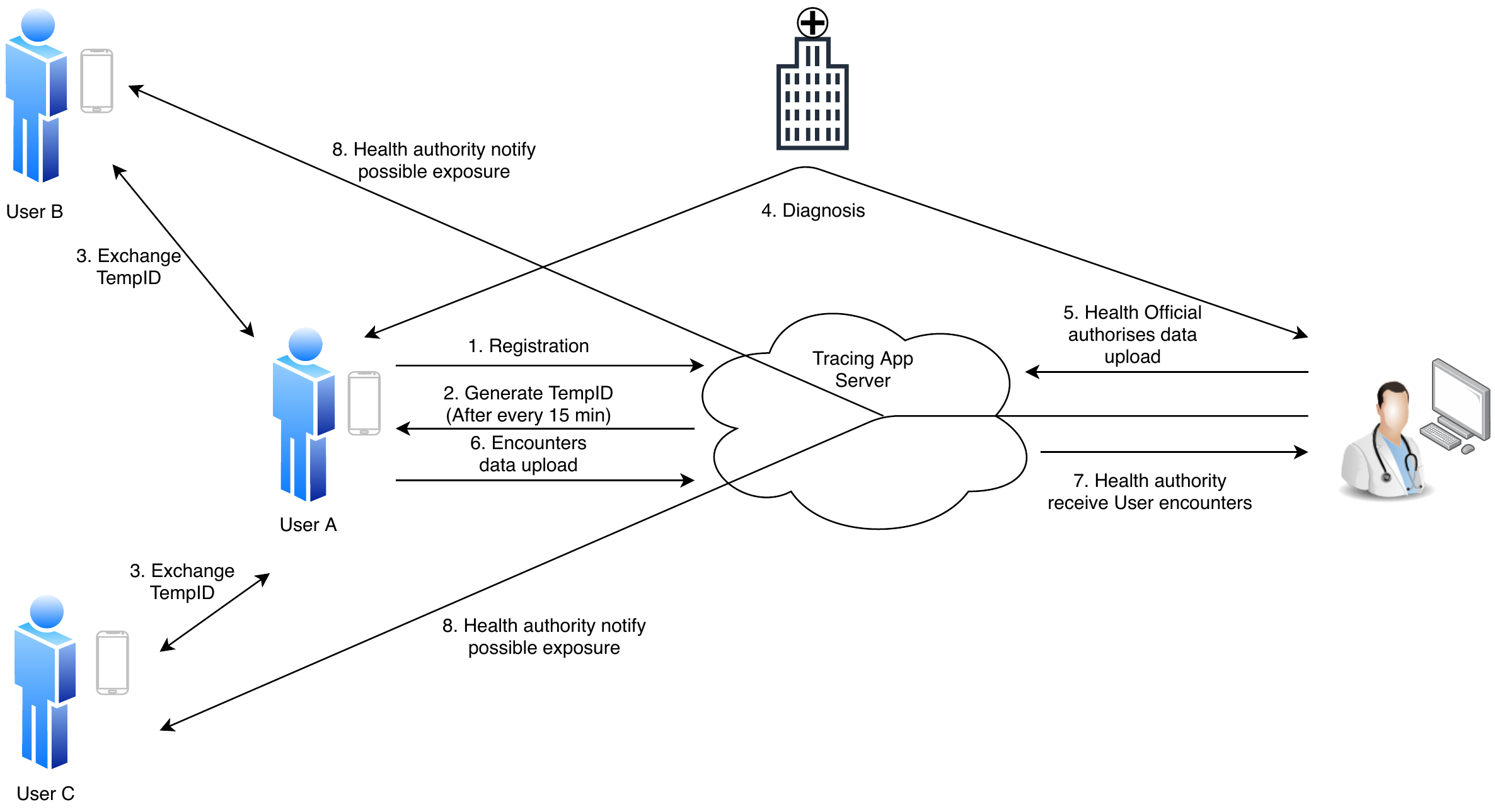}
    \caption{Tracing apps centralised architecture}
    \label{fig:mainArch}
\end{figure}

\section{System Architecture}
\label{sec:Arch}

The type of architecture adopted for  the  data collection aspects of tracing apps  has been a matter of much discussion due to both security and privacy concerns. We will discuss  three distinct system architectures commonly used or proposed for developing COVID-19 tracing applications. These are the \textbf{centralised}, the \textbf{decentralised}, and the \textbf{hybrid} approaches that combine features from both the centralised and the decentralised architectures.
Our classification criteria consider how the server is used and what data is required (or stored) by it. We now discuss each of the three architectures detailing their salient features. We will discuss some specific tracing apps that employ each of our three architectures in a later section.

\subsection{Centralised}
\label{sec:cen}
Figure \ref{fig:mainArch} shows the main entities and interactions of a centralised architecture. We note that the centralised architecture we describe is based on the Bluetrace protocol \cite{bluetraceWhitepaper}. The initial requirement for the app is that a user has to pre-register with the central server. The server generates a privacy-preserving Temporary ID (TempID) for each device. This TempID is then encrypted with a secret key (known only to the central server authority) and sent to the device. Devices exchange these TempIDs (in Bluetooth encounter messages) when they come in close contact with each other. Once a user tests positive, they can volunteer to upload all of their stored encounter messages to the central server. The server maps the TempIDs in these messages to individuals to identify at-risk contacts. More details on the centralised architecture's key processes are now given.

\subsubsection{Registration Phase}
\begin{figure}[htb]
    \centering
    \includegraphics[width=0.48\textwidth]{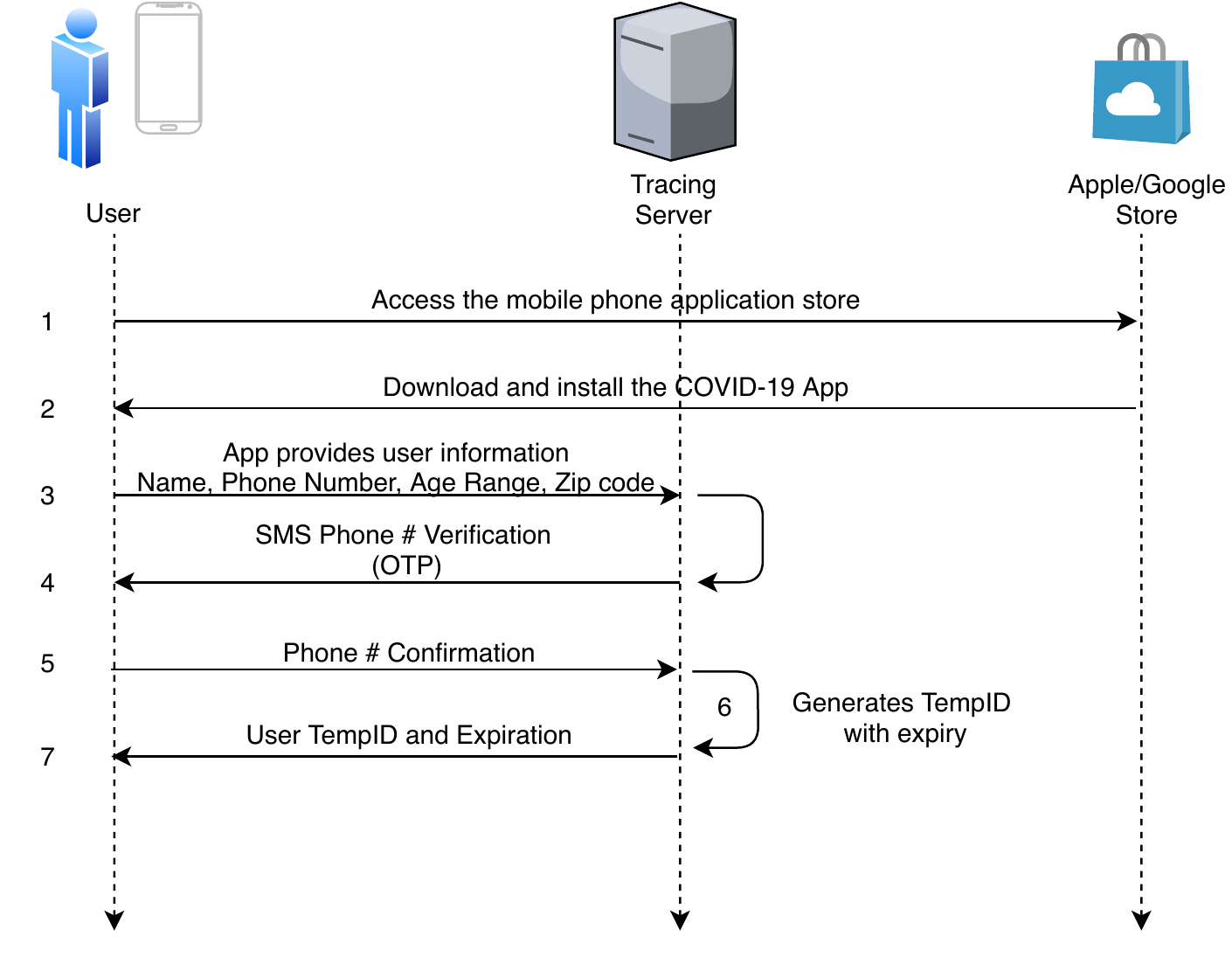}
    \caption{Centralised tracing app registration process}
    \label{fig:register}
\end{figure}

Figure~\ref{fig:register} shows the steps required to register a user in a centralised architecture. A user downloads the app (steps 1 and 2) and registers details such as name, mobile phone number, age bracket, and postcode with the server (step 3). The server verifies the mobile number by sending a One Time Password (OTP) by SMS (steps 4 and 5). Upon verification, the server computes a TempID (step 6), which is only valid for a short time (Bluetrace recommended expiry time is 15 min). The TempID and the expiry time are then transmitted to the user's app.

\subsubsection{Registering encounters/contacts information}\label{Cen-EncMessage}
\begin{figure}[htb]
    \centering
    \includegraphics[width=0.7\textwidth]{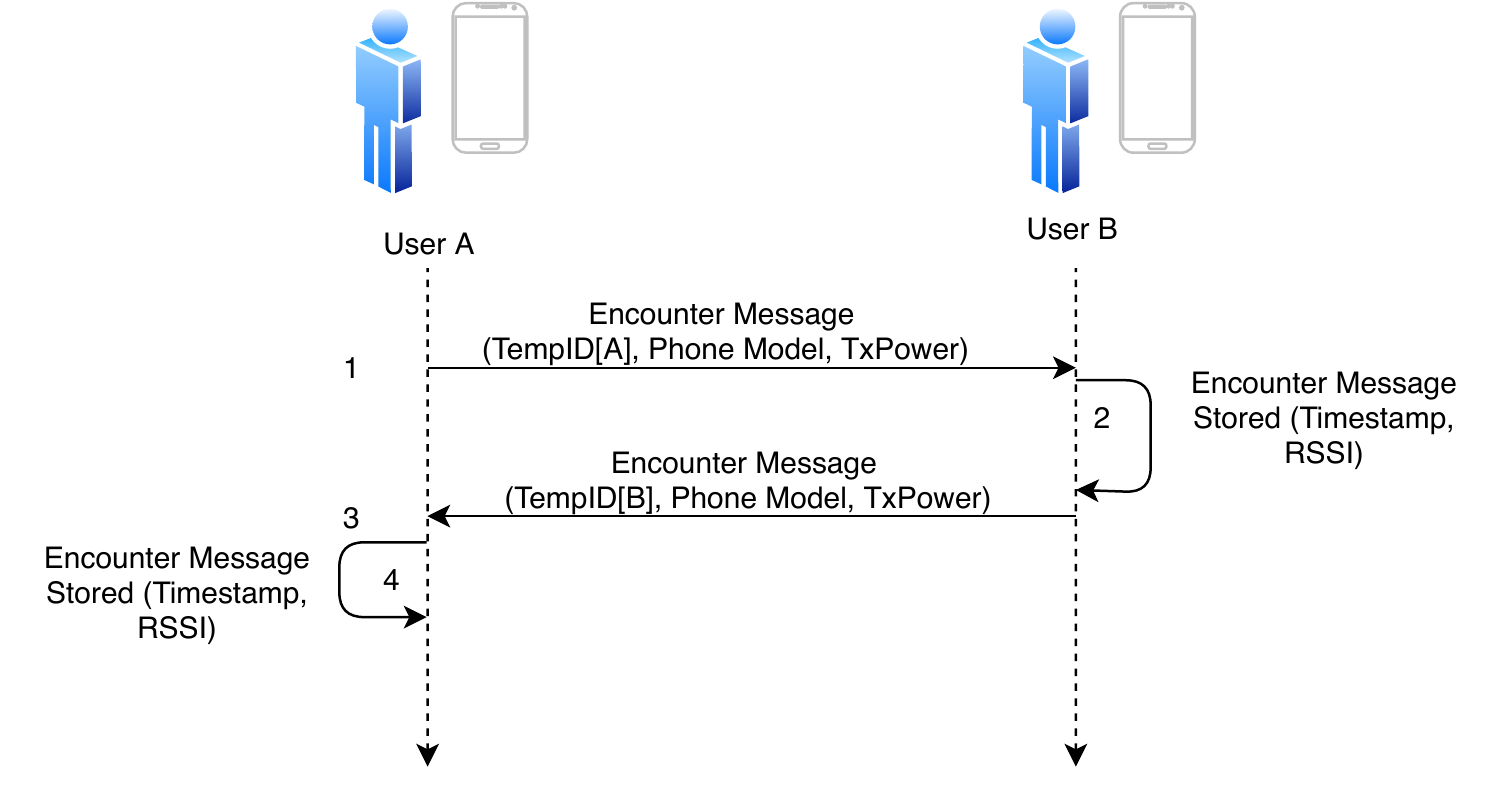}
    \caption{Centralised tracing app contact exchange operation}
    \label{fig:operation}
\end{figure}

Once a user comes in contact with another app user, they exchange an ``Encounter Message" using Bluetooth, as presented in Figure~\ref{fig:operation}. An encounter message comprises the exchange of TempID, Phone Model, and Transmit Power (TxPower)  (steps 1 and 3). Each device also records the Received Signal Strength Indicator (RSSI) and the timestamp of the message delivery (steps 2 and 4). Note that phone numbers are not included in these messages. Since the TempIDs are generated and encrypted  by the server they do not reveal any of the app user's personal information. Thus, both app users have a symmetric record of the encounter that is stored on their respective phones' local storage. The protocol uses a temporary blacklist to avoid a user registering duplicated contacts. Thus, once a user receives an Encounter Message, the app automatically blacklists the sender for a short time.


\subsubsection{Uploading encounters data}
\begin{figure}[htb]
    \centering
    \includegraphics[width=0.7\textwidth]{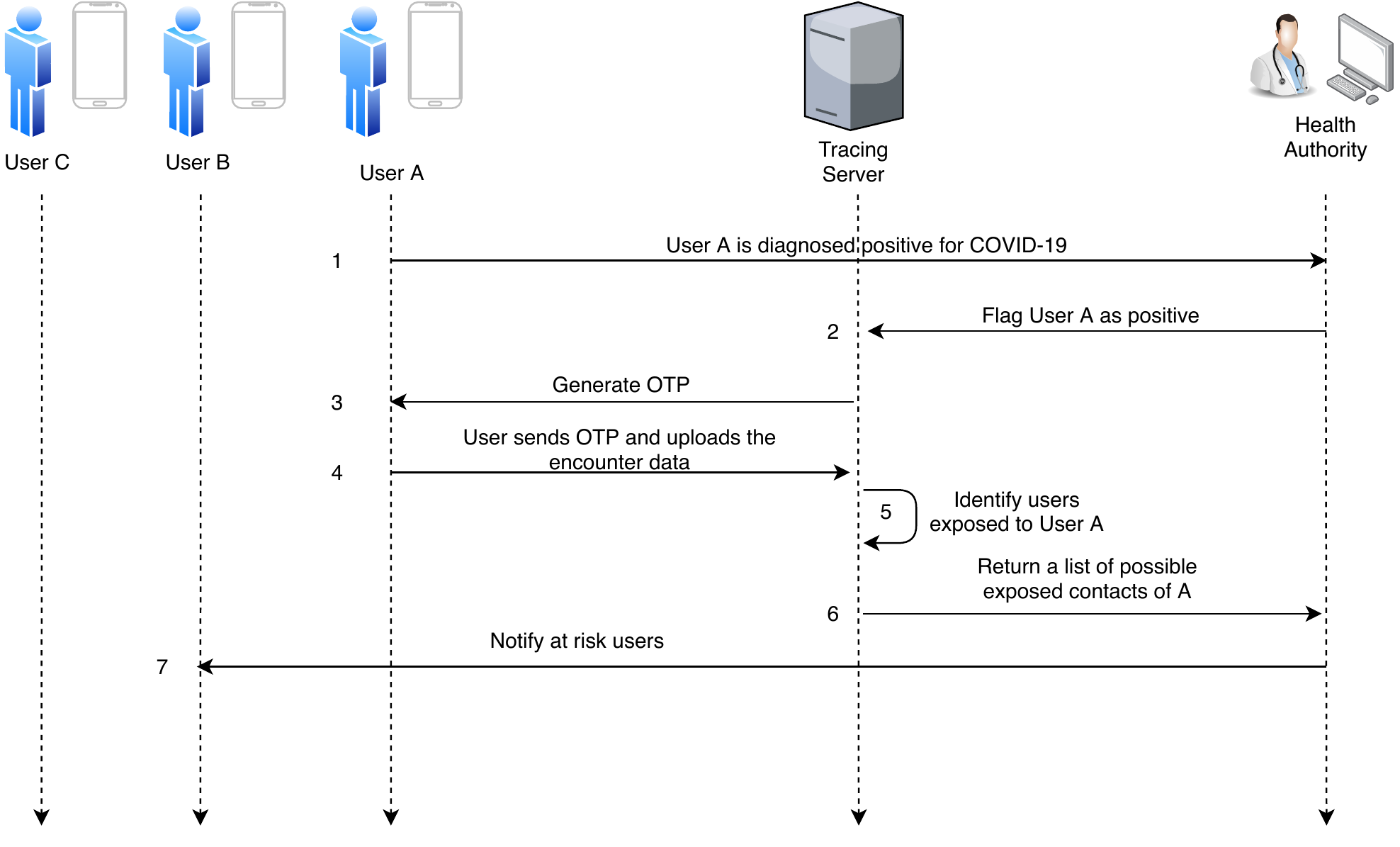}
    \caption{Centralised tracing app notification}
    \label{fig:notification}
\end{figure}

All encounter records are stored locally and are not automatically uploaded to the server. Figure~\ref{fig:notification} shows the application flow when a user tests positive for COVID-19 (step 1). The health official confirms whether the user has the tracing app installed, and flags the user as infected (step 2). The encounter data upload is voluntary. If the user agrees to upload the data, the health official sets this up in the back-end server, and the server generates an OTP for verification (step 3). Once verified, the encounter data is uploaded to the server (step 4).


\subsubsection{Server-side processing of the uploaded data}
The server iterates through the list of encounter messages, decrypting each TempID with its secret key. This TempID is then mapped to the user's mobile number. The server uses the TxPower and RSSI values to approximate the distance (proximity) separating the users during the reported encounter. The proximity estimation can also be performed locally on the phone, but this has battery usage implications. This proximity data, in conjunction with the timestamps, is used to ascertain the risk profile (closeness and duration) of the encounter (step 5, Figure \ref{fig:notification}). A list is prepared with all the required information (step 6) for further processing by the relevant health official (step 7).


\begin{figure*}[htb]
    \centering
    \includegraphics[width=0.9\textwidth]{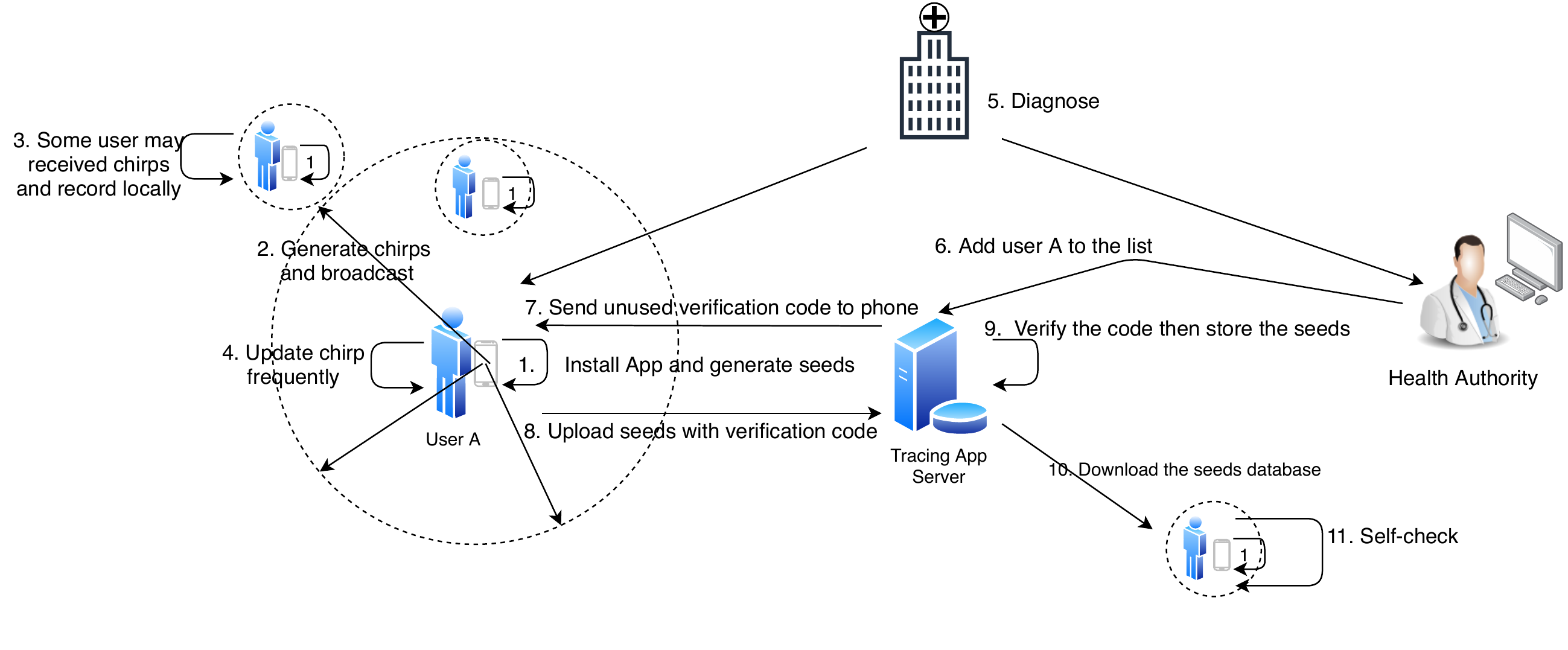}
    \caption{Tracing apps decentralised architecture.}
    \label{fig:de-overview}
\end{figure*}

To summarise: In the centralised architecture, the central server plays a key role in performing core functionalities such as storing encrypted PII information, generating anonymous TempIDs, risk analysis, and notifications for close contacts. This accumulation of responsibilities raises privacy concerns that are discussed in detail in Section \ref{sec:Privacy}. The server is assumed trusted in this architecture, with some countries introducing strict privacy-protection regulations for safeguarding the use and life cycle of the collected data \cite{regulation1}.

\subsection{Decentralised}
\label{sec:dec}
In contrast to the centralised architecture, the decentralised architecture proposes to move  core functionalities to the user devices, leaving the server with minimal involvement in the contact tracing process. The idea is to enhance user privacy by generating anonymous identifiers at the user devices (keeping real user identities secret from the other users as well as the server) and processing the exposure notifications on individual devices instead of the centralised server. We discuss the privacy and security implications of this design in Section \ref{sec:Privacy}.

We take the Private Automated Contact Tracing protocol (PACT) \cite{pact-ec} as a base to describe the decentralised architecture. The decentralised approach does not require app users to `pre-register' before use, thus avoiding the storage of any PII with the server.  Devices generate their random seeds (used as input for a pseudorandom function), which are used in combination with the current time to generate privacy-preserving pseudonyms or `chirps' with a very short lifetime of about 1 min (see Figure~\ref{fig:de-overview}). These chirps are subsequently periodically exchanged with other devices that come in close contact. Once a user is positively diagnosed with COVID-19, they can volunteer to upload their seeds and the relevant time information  to a central server.
This is in contrast to the centralised architecture where the complete list of encounter messages is uploaded. Uploading of seeds, instead of all used chirps, improves latency and provides improved bandwidth utilisation.

The central server only acts as a rendezvous point, akin to a bulletin board to advertise the seeds of the infected users. This server is considered `honest-but-curious'. Other app users can download these seeds to reconstruct the chirps (by using timestamps) that were sent by the infected users. The server, as well as other users, cannot derive any identifying details just by knowing the seeds and chirps. Only the other app users can perform a risk analysis to check if they are exposed for a long enough duration. This one-way lookup against the downloaded seeds restricts the server's functionality and alleviates some of the privacy risks (see Section \ref{sec:Privacy}). More details on the decentralised architecture's key processes are now given.

\subsubsection{App Installation}
\begin{figure}[htb]
    \centering
    \includegraphics[width=0.7\textwidth]{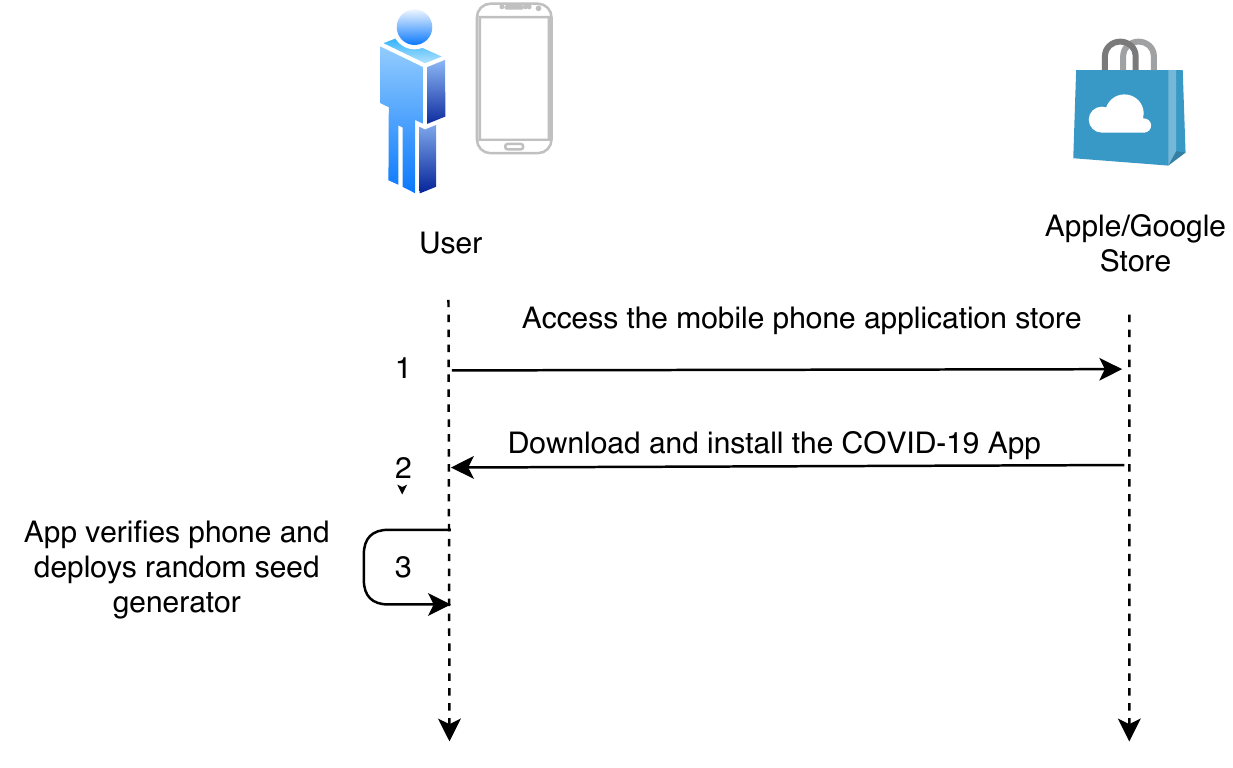}
    \caption{Decentralised tracing app installation process}
    \label{fig:de-install}
\end{figure}

COVID-19 tracing apps that adopt the decentralised architecture do not necessarily require an interactive registration process during the app installation stage. The app installation process only verifies a user's smartphone and deploys a random seed generation algorithm that is not linked to the phone (see Figure~\ref{fig:de-install}).

\subsubsection{Generating seeds, chirps and exchanging chirps}
\label{De-Chirps}
\begin{figure}[htb]
    \centering
    \includegraphics[width=0.7\textwidth]{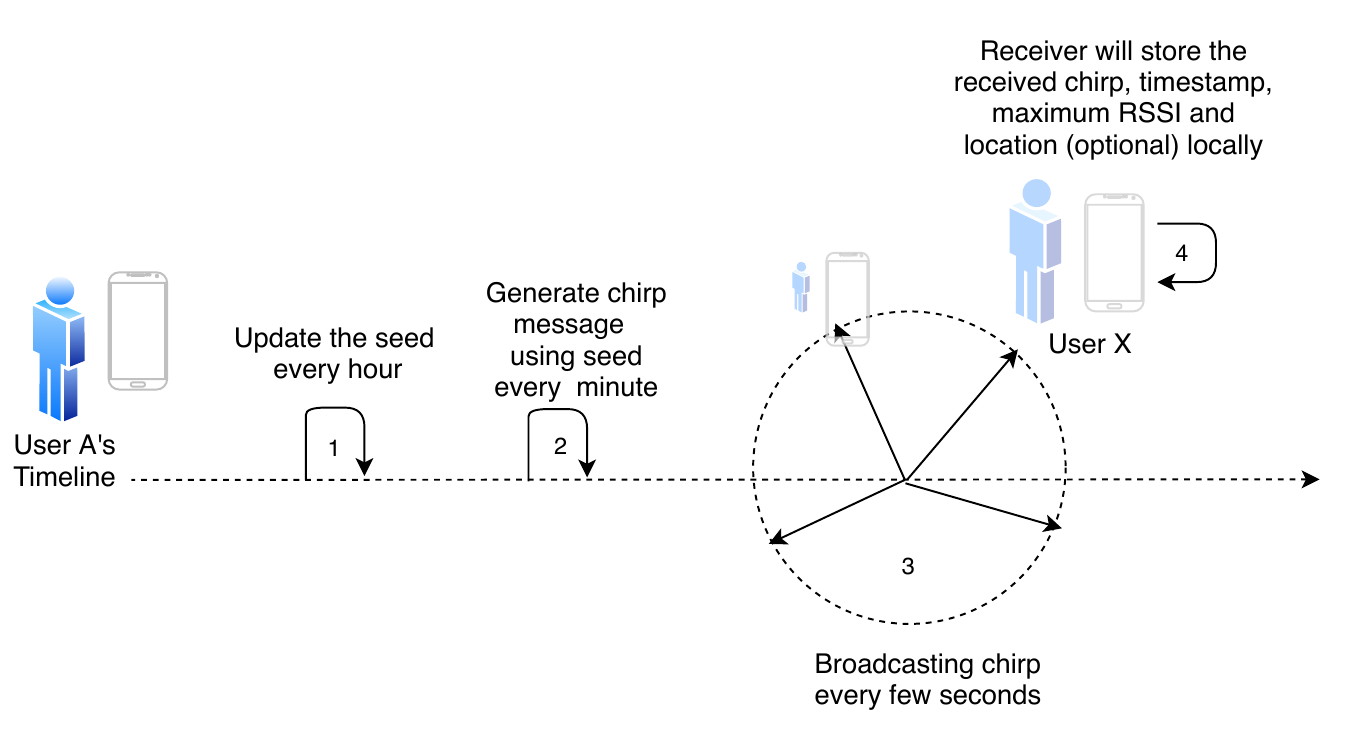}
    \caption{Decentralised tracing app encounter exchange}
    \label{fig:de-encounter}
\end{figure}

Once the decentralised tracing app is installed, the seed is generated (with an expiry period of one hour) by the user's device (see Figure~\ref{fig:de-encounter}). This seed and the current time are subsequently used in a pseudorandom function to generate the chirp.
The chirps are not linked to an individual or their phone - so in principle, they are anonymous. The app generates new chirps with a time granularity of 1 min.
These are broadcasted every few seconds via the Bluetooth beacon. In the listener's phone, the app will automatically store all chirps received (step 4 in Figure~\ref{fig:de-encounter} ). The  information stored in the receiving app includes the chirp, the timestamp when the chirp is received, and the maximum RSSI value. Identical chirps received within 1 min are ignored.
Note the critical difference from the centralised architecture where TempIDs are created by the server - in the decentralised case, the seeds and chirps are generated at the device.


\subsubsection{Uploading encounters data}
\label{De-upload}
\begin{figure}[htb]
    \centering
    \includegraphics[width=0.7\textwidth]{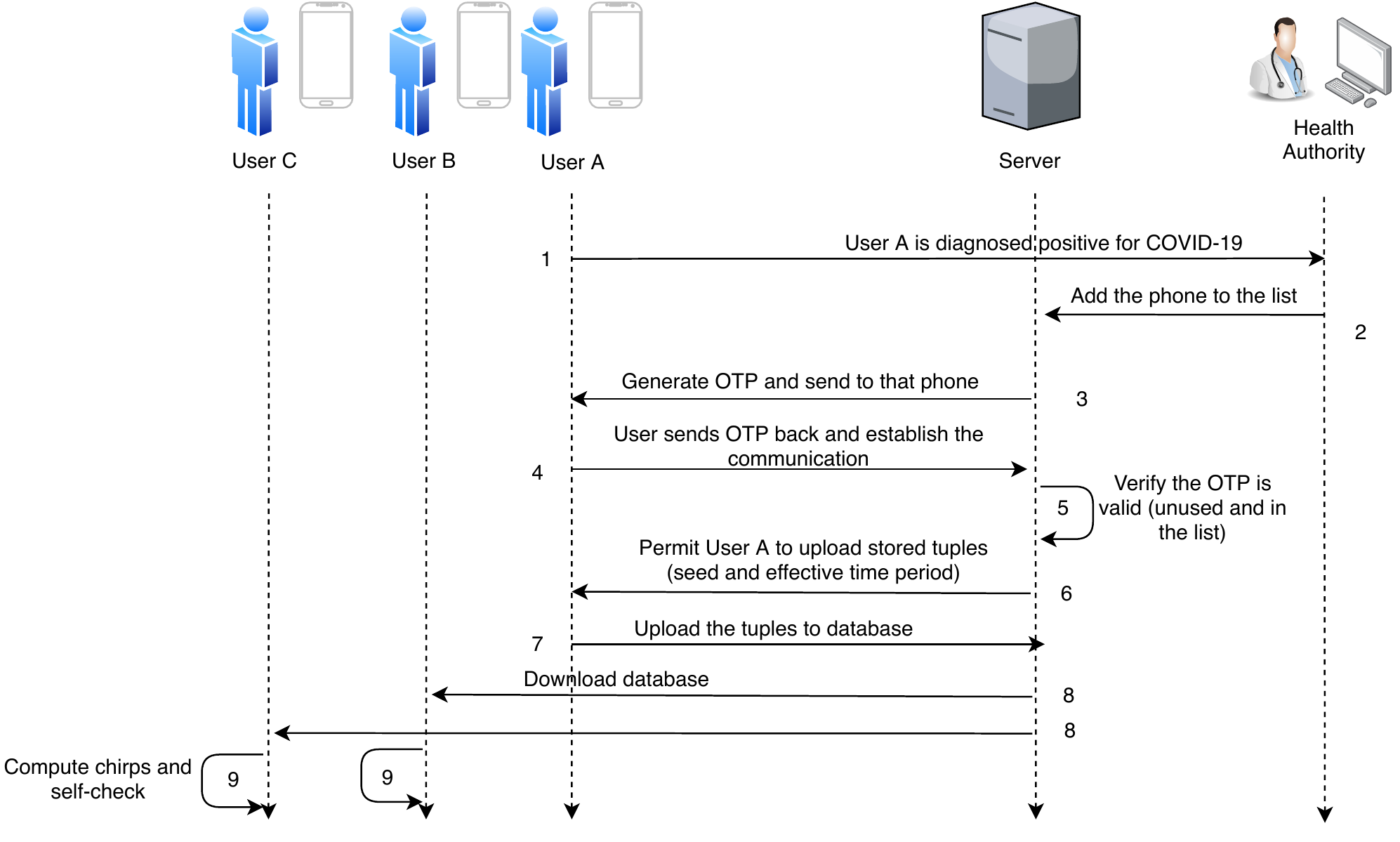}
    \caption{Decentralised tracing app tracing process}
    \label{fig:de-trace}
\end{figure}

If a user is diagnosed positive, they are given a unique ``permission number" by the relevant authority to authorise the upload of all used seeds that are locally stored in their phone (illustrated in Figure~\ref{fig:de-trace}), as well as the creation and expiry times of the seeds. Note, the server in the decentralised architecture only gets the seeds associated with a single identified user. This is to be compared with  the centralised architecture where the complete contact list (with TempIDs) of all encountered individuals is uploaded to the server.

\begin{figure*}[htb]
    \centering
    \includegraphics[width=0.9\textwidth]{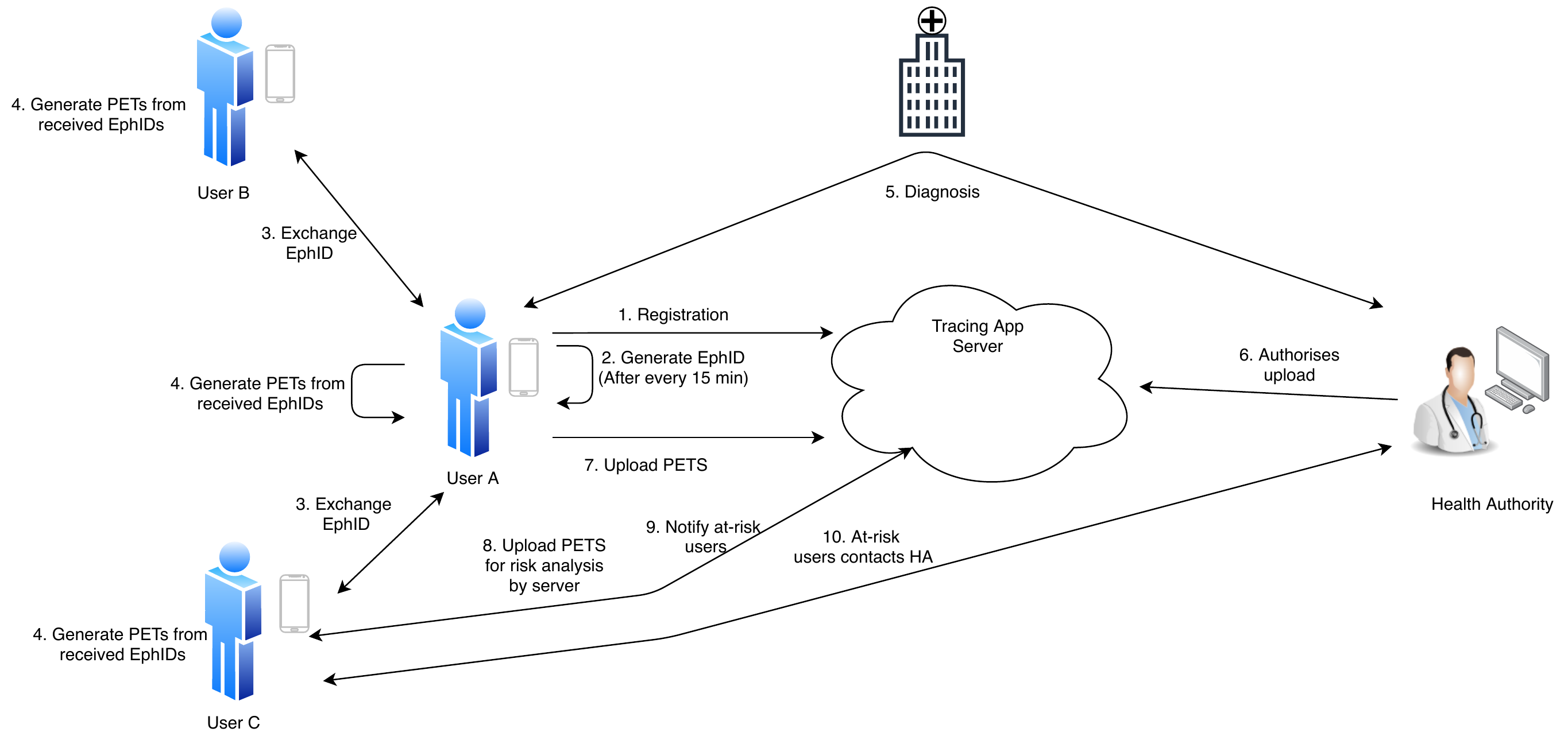}
    \caption{Hybrid tracing app architecture}
    \label{fig:hybrid}
\end{figure*}

\subsubsection{The contact tracing process}
\label{De-tracing}
Contrary to the centralised architecture, the tracing process in the decentralised architecture is performed locally by the app user on their device (instead of the central server).  The app users can communicate with the server, typically once per day, to download any seeds uploaded by infected users. Given such seeds are downloaded (step 8 in Figure~\ref{fig:de-trace}), the user's app then reconstructs all the corresponding chirps (using pseudorandom calculations based on the seeds and discrete-time intervals between the start and expiry time). Finally, the app performs a lookup to check if any of the reconstructed chirp information appears in its local encounter chirp log. If so, proximity and duration times are then derived (based on timestamps and RSSI values) for risk analysis purposes. No human intervention is required.

\subsection{Hybrid}
\label{sec:hybrid}
In the centralised architecture, the server performs all the complex tasks, e.g., TempID calculations, encryption, decryption, risk analysis, and notifications of alerts for the at-risk contacts. On the other hand, all these functionalities are delegated to devices in the decentralised architecture, keeping the server only as a bulletin board for lookup purposes. The hybrid architecture proposes that these functionalities are split between the server and the devices. More specifically, the TempID generation and management remain decentralised (i.e.,  handled by devices) to ensure privacy and anonymisation, whilst the risk analysis and notifications should be the responsibility of the centralised server. There are three main reasons for performing the tracing process at the server: \textit{i}) In the decentralised architecture, the server is unaware of the number of at-risk users as the devices make this risk analysis without taking the server into consideration. Thus, the server does not have any statistical information and is unable to run any data analytics to identify exposure clusters. \textit{ii}) Risk analysis and notifications are considered a sensitive process that should be handled by the authorities, keeping the existing infrastructure resources and state of the pandemic in mind. \textit{iii}) The uploaded encounter information from infected users is not made available to the other users but retained only at the server. This is to avoid user de-anonymisation attacks (details in Section \ref{sec:linkage}) possible in the decentralised architecture.

Figure \ref{fig:hybrid} shows the interaction sequence in the hybrid architecture based on the Desire protocol \cite{Desire}. This protocol requires the user's app registration process to assign a unique device ID without recording any PII. Devices then cryptographically generate and exchange Ephemeral IDs with other devices over BLE. For each received EphID, two un-linkable Private Encounter Tokens (PETs) are generated and stored to represent an encounter. Once a user has tested positive, a list of the locally generated PETs is uploaded to the server. Any device can now send their second generated PET tokens to the server, which then performs risk analysis and notification. The server cannot infer any identifying information from the PETs, and all communication between the server and devices is routed through a proxy or anonymisation  network. More details on the hybrid architecture's key processes are now given.
\begin{figure}[htb]
    \centering
    \includegraphics[width=0.48\textwidth]{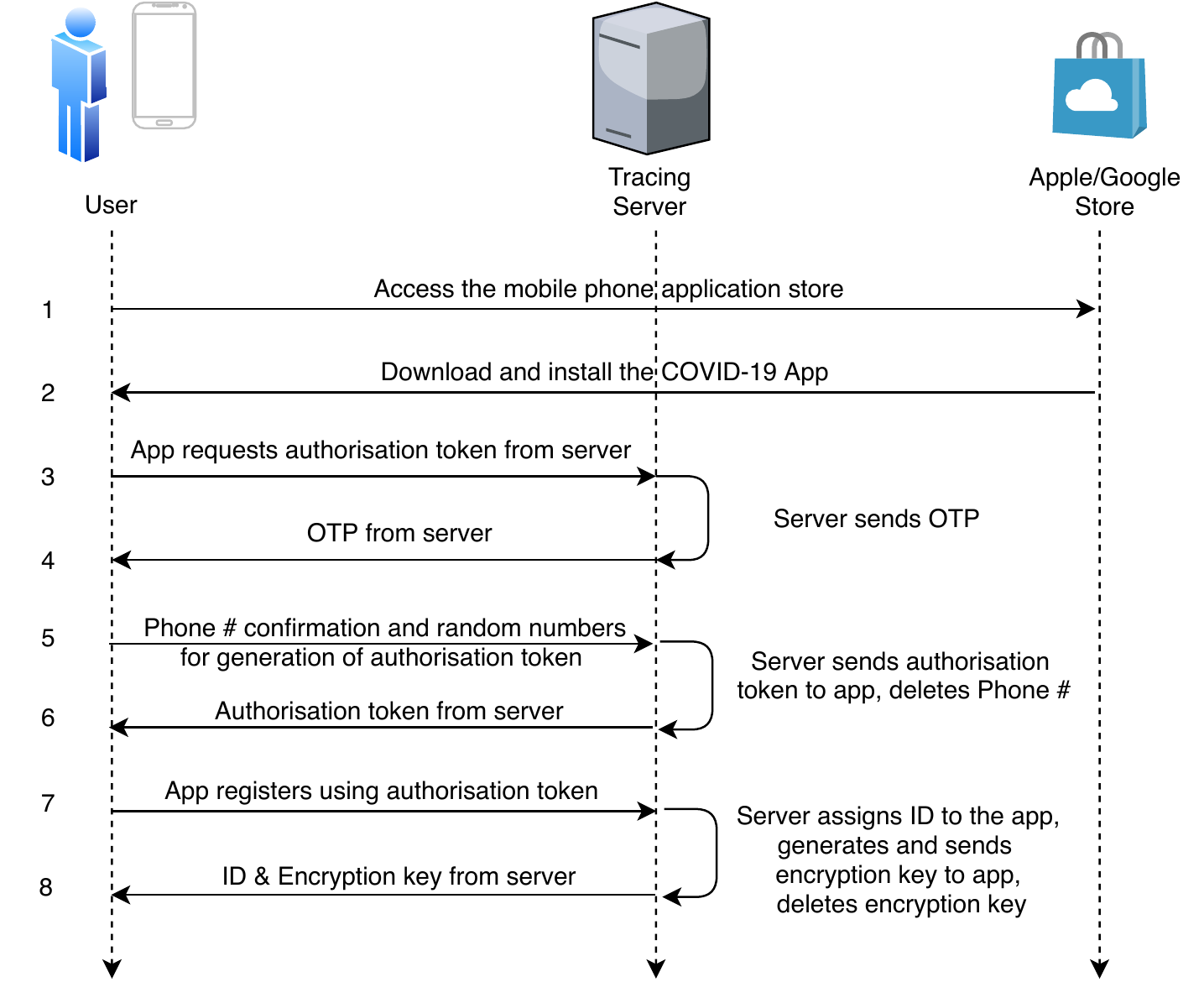}
    \caption{Hybrid tracing app registration process}
    \label{fig:hybridRegister}
\end{figure}

\subsubsection{Installation and registration}
The registration process in the hybrid architecture requires a two-step authentication process, wherein the phone number is verified by the OTP, and the app is verified through an authorisation token issued by the server. Figure \ref{fig:hybridRegister} shows the process. As the server is not allowed to store any PII, it deletes the phone number after verification. The server then assigns the app a unique ID and generates an encryption key that is sent to the app. The server then deletes the encryption key. The client, in the future, will identify themself using this ID.

\subsubsection{Generating and exchanging Ephemeral IDs}

\begin{figure}[htb]
    \centering
    \includegraphics[width=0.8\textwidth]{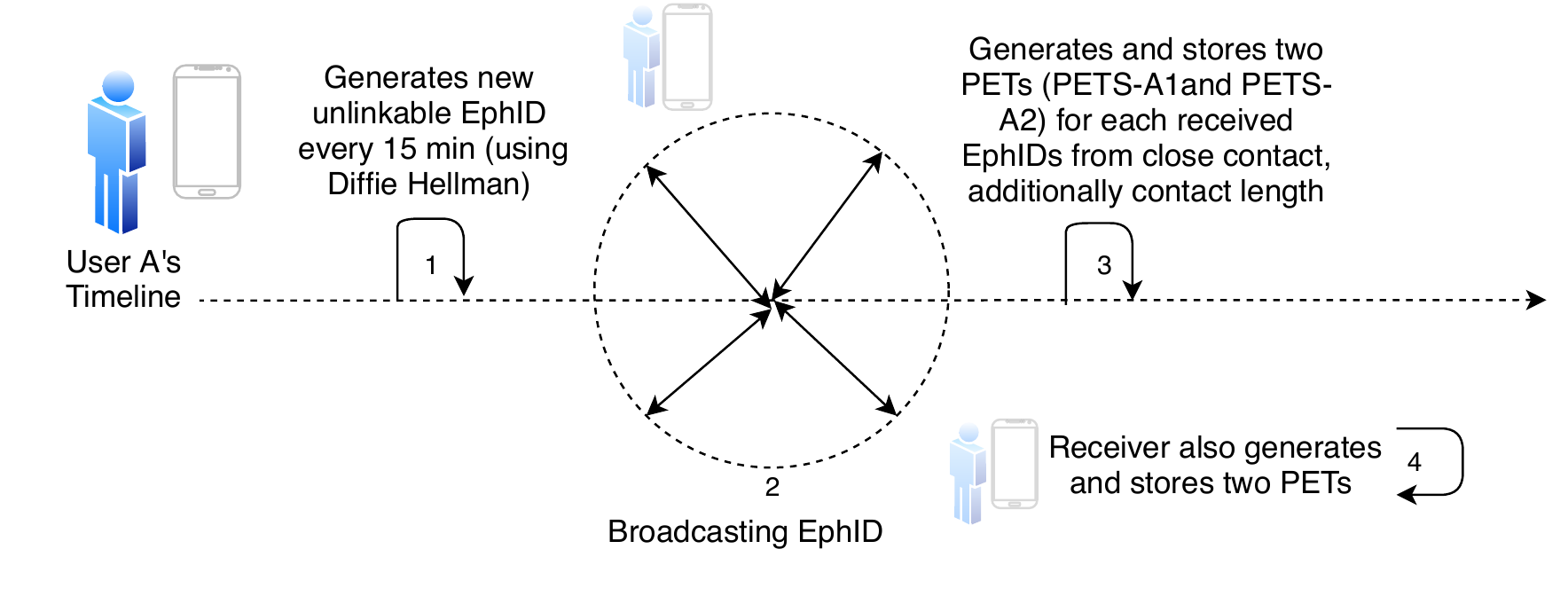}
    \caption{Hybrid tracing app encounter process}
    \label{fig:hybridEncounter}
\end{figure}

In the operation phase, the device generates a new Ephemeral ID (EphID) using the Diffie Hellman key exchange mechanism \cite{DH}, which is valid for typically 15 min and synchronised with the Bluetooth MAC address rotation interval. The device starts broadcasting this EphID (= $g^a$) through Bluetooth. Once an EphID is received from another device (say with exponent b), the app generates two PETs \{PET1 = $H('1'| g^{a.b})$, PET2 = $H('2'| g^{a.b})\}$. The app maintains two tables, referred to as upload and query tables. One PET is stored in the query table, and the other one, along with the time and duration of the contact, is stored in the upload table.

\begin{figure}[htb]
    \centering
    \includegraphics[width=0.8\textwidth]{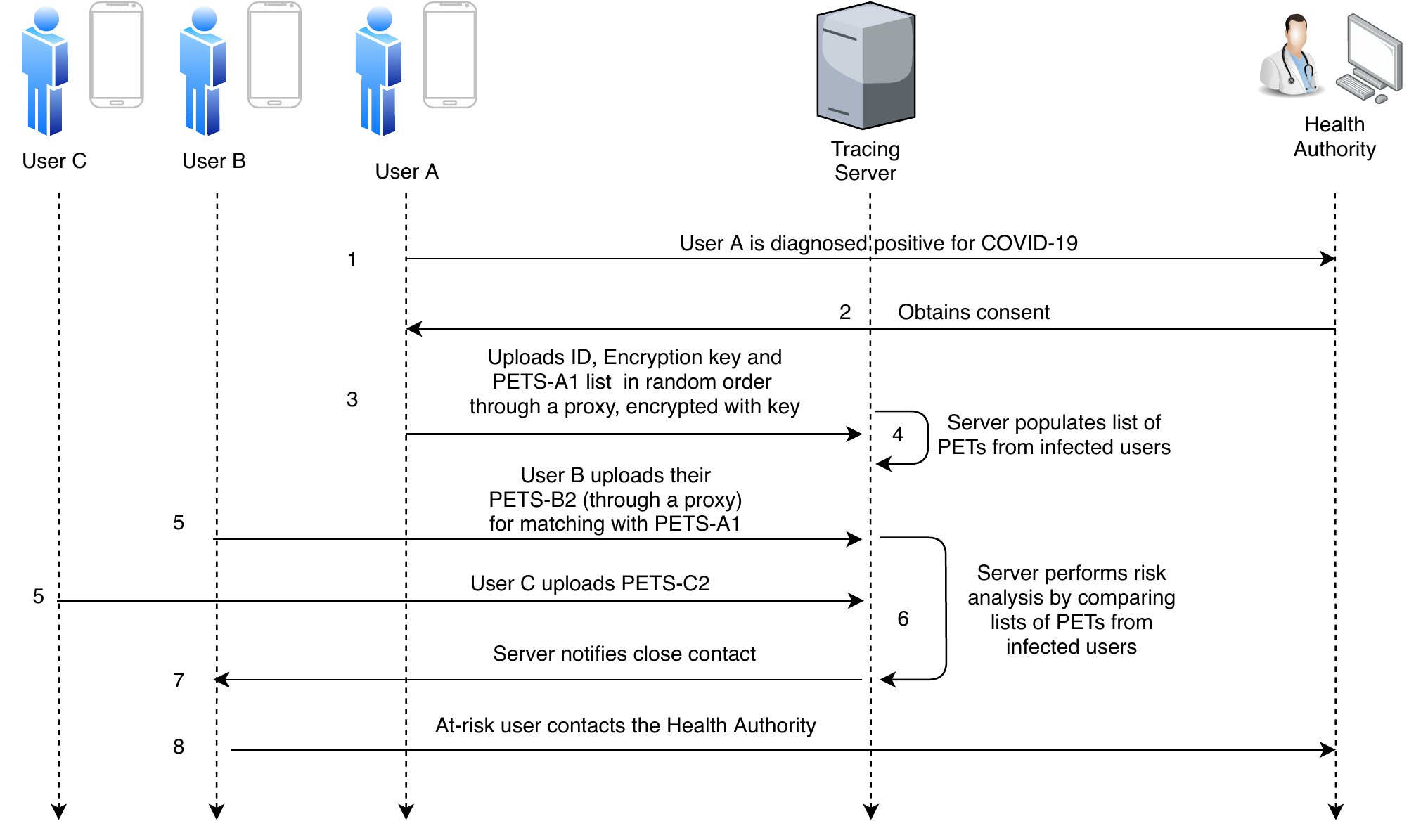}
    \caption{Hybrid tracing app notification process}
    \label{fig:hybridNotification}
\end{figure}

\subsubsection{Uploading Encounter data}
Once a user is diagnosed with COVID-19, explicit consent is required for the data upload. The user uploads ID, encryption key, and PETs in the upload table along with time and duration values (steps 1-3 Figure \ref{fig:hybridNotification}). The server records these PET values and associated data and uses the encryption key to update the user record (status).

\subsubsection{Contact tracing process}
Any user who wants to check their risk exposure to an infected case uploads the PETs in the query table to the server by using a proxy or anonymisation network (step 5 Figure \ref{fig:hybridNotification}). The server performs risk analysis by matching the PETs from the query table with the PETs uploaded by the infected user. Using the time and duration values, the server evaluates whether the user is at-risk or not. Any at-risk user is notified through the app to contact the health authority. Note that the server is not able to identify any user based solely on their PET values.

\subsection{Architecture summary}
We have discussed the three base architectures employed for developing applications for contact tracing purposes. The architectures are categorised based on the functionality and level of privacy preservation at the central server. In the centralised architecture, the server manages the security keys, generation of anonymous IDs, contact risk analysis, and notification processes. All these roles are transferred to the devices in the decentralised architecture while the server acts simply as a bulletin board.  The hybrid architecture tries to balance the load on the server and improve privacy preservation by splitting functionalities between the end-user device and the server.
One distinct advantage of using an architecture that pushes the risk analysis and notification process to the centralised server (i.e., both centralised and hybrid architectures) is that  health officials can decide the rate of notifications depending on the pandemic circumstances (e.g., the availability of test kits).  On the other hand, decentralised and hybrid architectures aim to keep the user identities secret from the central server. A server security breach in this latter architecture would, therefore, result in lower  information leakage.

\section{Data Management, Privacy and Security}
\label{sec:Privacy}

One of the major issues in any tracing application is management, privacy, and security of the data that is collected. The European Data Protection Board issued a statement on the importance of protecting personal data while fighting COVID-19 and flagged articles of the General Data Protection Regulation that provide the legal grounds for processing personal data in the context of epidemics \cite{aruna1}. In addition, some Governments' have passed special privacy protections laws aimed at addressing privacy issues \cite{aruna2}. To comply with these requirements, the tracing apps need to use a multitude of techniques, across the three distinct phases of their operation: i) Registration, ii) Operation, and iii) Positive case identification phases, depending on:
\begin{itemize}
    \item What data is produced and by whom?
    \item What data is exchanged between whom and when?
    \item What data is stored where and by whom?
    \item Who can access what piece of data?
\end{itemize}

In this section, we will first discuss the data life cycle for the three architectures described in Section \ref{sec:Arch} and then focus on the privacy and security issues associated with these architectures. We consider three key stakeholders in the attack ecosystem; i) the government ii) the administrator controlling the central server (referred to as server for brevity), and iii) malicious users. All architectures assume that health authorities\footnote{Health authorities include health officials, testing centres, and allied facilities managing COVID-19. 
}
already know the real identities of all positive cases as all uploads are authorised through the health authorities to prevent fake data being uploaded.
\begin{table*}[ht]
  \footnotesize
  \centering
  \caption{Data storage for centralised tracing architecture}
  \label{table:DLC-C}
    \begin{tabular}{ |c|l|l|l| }
      \hline
      Storage & Registration phase & Operation phase & Positive case identification phase
      \\
      \hline
      Server & Phone Number, Name, & Generates and stores  & List and contact details of all
      \\
      & Age range, ZIP code & TempIDs for each user & i) positive cases
      \\
      & & &ii) close contacts of each positive case
      \\
      \hline
      Devices & - & Own TempID \& Phone model& -
      \\
      &  & Encounter messages received from contacts&
      \\
      &  &(TempIDs of contact, time stamp, RSSI &
      \\
      &  &TxPower, Phone model) &
      \\
      \hline
    \end{tabular}
  \end{table*}

\begin{table*}[ht]
  \footnotesize
  \centering
  \caption{Data storage for decentralised tracing architecture}
  \label{table:DLC-D}
    \begin{tabular}{ |c|l|l|l| }
      \hline
      Storage & Registration phase & Operation phase & Positive case identification phase
      \\
      \hline
      Server & - & -  & Seeds (and validity period) received from all positive cases
      \\
      &  &  &
      \\
      \hline
      Devices & -  & Generates own seeds and chirps& Seeds for all positive cases received from the server
      \\
      &  &  &
      \\
      &  &Chirps received from contacts,& Generates chirps based on seeds/validity period of positive cases
      \\
      &  &time stamp, RSSI  &
      \\
      \hline
    \end{tabular}
  \end{table*}

\subsection{Data management}
\label{sec:data}
Details of data storage in the centralised architectures appear in Table \ref{table:DLC-C}. The server is responsible for i) storing PII collected when a user registers, ii) generating,  storing and transferring the TempIDs to all registered users periodically (after every 15 min for example), and iii) maintaining a list of all individuals who are diagnosed as positive and their close contacts. In contrast, the user device,  after receiving the TempID from the server, carries out the following two tasks:  i) it generates, exchanges and stores the contacts it has had  with peers for a specified period of time, usually 21 days, and ii) upon request, it shares contact data it has stored  with the server with the consent of the user.

Data storage details for the decentralised architectures are presented in Table \ref{table:DLC-D}. The user devices are responsible for generating the hourly seed and computing the chirps based on the seed and the current time. In addition, they are responsible for  exchanging and storing these chirps, the RSSI, and the received timestamp information with peers. There is also the option for the device to store additional metadata, such as location information.  The server plays a limited role compared to the centralised architecture. It is only called into action when a user is diagnosed as COVID-19 positive and voluntarily uploads the seeds and time validity data as described in Section ~\ref{De-Chirps}. This data,  stored at the server, can now be used for lookup by other users who have come in contact with the infected user, by reconstructing the chirps using the seeds.

\begin{table*}[ht]
  \footnotesize
  \centering
  \caption{Data storage for hybrid tracing architecture}
  \label{table:DLC-H}
    \begin{tabular}{ |c|l|l|l| }
      \hline
      Storage & Registration phase & Operation phase & Positive case identification phase
      \\
      \hline
      Server & Device ID & Device ID  & Stores PETs (and validity period) received from all positive cases
      \\
      &  &  & Stores metadata about positive cases
      \\
      &  &  & Stores query PETs from other users
      \\
      \hline
      Devices & Device ID, & Generates and store own EphID & -
      \\
      &  &  &
      \\
      & Encryption key & Maintain two tables of PETs &
      \\
      &  &Stores timestamp, duration &
      \\
      &  &and signal strength for PETs &
      \\

      \hline
    \end{tabular}
  \end{table*}

Table \ref{table:DLC-H} shows data storage during the various phases of  hybrid tracing architecture. In the operation phase, the devices store all encounters as PET entries in two different tables.  The server records the device IDs (with blank metadata such as risk score, notified or not, etc.). The server only obtains the PETs from users who have tested positive and volunteers to upload this information. Another significant difference from the decentralised architecture is that these PETs are not transferred to other devices; rather, other devices upload their PETs from their query table for a risk analysis to be carried out  by the server.

\subsection{Privacy}
The success of any automatic contact tracing app depends on several factors, including: how seamlessly and accurately it can capture close contacts. Another factor is the confidence the users have about  their privacy and security when using the app. A naive approach for contact tracing could be to develop a privacy-agnostic system that advertises and exchanges the mobile phone numbers of the participants and periodically registers their location with a centralised server \cite{pact-ec}. Such an application would raise serious privacy concerns, and would likely not be accepted by users. Therefore, all the architectures have  privacy protection built-in. However, the amount of protection provided differs considerably and depends on the attack models, trust assumptions, and the protection measures they adopt \cite{Fraunhofer}.


In the previous section, we discussed the data management aspects highlighting the source and storage of different types of data in the three architectures. From a privacy perspective, we classify the data that is to be stored into three categories: i) PII of  participants (e.g., names, phone numbers, whether they have tested positive to the virus or not, etc.), ii) Contact advertisement messages (pseudonyms exchanged between devices), and iii) Social/proximity graphs; an indication of the interactions between users and the people they came into close contact with. Each data category has different privacy implications.

We first explore the smartphone's privacy implications, as it is typically less secure than a server. In this case, attacks like theft or coercion (a user being forced or persuaded) will result in the content stored in the smartphone being revealed. This type of threat is present in all  of the architectures. However, the difference between the different  architectures is what is stored on the smartphones (see Tables \ref{table:DLC-C}, \ref{table:DLC-D}, \ref{table:DLC-H}). Data that may be stored on devices, such as details of encounter messages, is considered to be less sensitive, as this information cannot be used to directly identify the contacts.

In a centralised architecture, the servers have access to all three types of data. Therefore, if access to the servers is compromised by malicious users, it would be possible to identify all individuals and their contacts, therefore jeopardising their privacy.
Hence, centralised architectures need to provide adequate protection of the  servers to guarantee user privacy. 

In the decentralised architecture, all users can access the public server to download the list of seeds and calculate the chirps used by an infected user. However, as these seeds are uploaded together with their expiry periods, they can result in the unauthorised identification of  infected individuals using other side-channel information. For example, malicious persons/apps/organisations  can keep collecting the ephemeral identifiers and the seeds from reported cases and link the identifiers/chirps with the accessed auxiliary information. Also, as only an infected user uploads seeds to the server, a traffic analysis attack, launched by a malicious user who can eavesdrop, would be able to identify a COVID-19 positive user uploading seeds to the server. These attacks are discussed further in section \ref{sec:Attacks}.

The hybrid architecture adopts additional advanced privacy enhancement methods such as secret sharing \cite{secretSharing}, decisional Diffie-Hellman (DDH) \cite{ddh}, and private set intersection \cite{PSI}\footnote{Readers are encouraged to follow the provided references for details of these privacy preserving, secret sharing techniques.}.  In general, a  user's secret is shared by the user and the server. Furthermore, part of the infection risk analysis is computed at the server using privacy preserving secret sharing. Therefore, if one party is compromised, the entire secret or risk analysis result will not be revealed. These privacy enhancement methods help protect the identity of infected users from being revealed by malicious users or compromised servers. However, these enhanced privacy protections still cannot prevent the users' PII getting de-anonymised if a malicious user can successfully access data collected from side-channel context information.


\subsection{Security}
The notion of security encompasses limiting an adversary's abilities to introduce false negatives and false positives in the system, in addition to ensuring system integrity and availability. The motivation for carrying out an attack varies and can range from political and ideological to financial. In the context of contact tracing, an attacker may aim to inject erroneous entries or cause a denial of service.


As all three architectures discussed in this article involve a centralised server, it is pertinent to explore the specific security threats for each of the architectures. The potential security threat depends on what data originates from a server,  what data is shared and accessible to a server and in what form the data is collected and stored (e.g., pseudonymous, encrypted, unencrypted). Furthermore, it depends on the modus operandi of the  server, namely whether it is  i) A trusted server, ii) An honest-but-curious server, iii) A compromised/malicious server, or iv) A colluding server.

 A malicious/compromised server can disrupt all types of communications or inject false exposure notifications in all architectures. Similarly, a colluding server can liaise with other malicious entities to perform user de-anonymisation.

In the centralised architecture, the server is considered trusted. It is responsible for storing users' PII and managing security keys used to encrypt/decrypt TempIDs. This poses the risk of data theft if the server gets compromised, a general threat against any centralised server. In this context, the server application needs to run in a trusted environment and use appropriate authentication and access control mechanisms. All information exchanged between the server and the user's smartphone as well as between the server and the health officials needs to be authorised and secure. Thus, centralised architectures only consider malicious users in their attack models and aim to keep the information of all users secure to prevent loss of users' privacy as described in Section \ref{sec:linkage}. This ensures that no malicious third party can access any information sent/received or exfiltrate information. However, malicious users in  centralised architectures could exploit the un-authenticated BLE contact information exchanged between devices to spread incorrect contact information by relaying or replaying. This type of attack,  discussed further in Section \ref{sec:replay}, would result in false positives during the contact tracing process, forcing users to be incorrectly notified as close contacts.

Decentralised and hybrid architectures, on the other hand, assume an honest-but-curious server that performs all the tasks assigned to it and passively harvests sensitive data, if available. The attack model considers the government and the server to be untrustworthy and only reveals users' identities to the health authorities. As mentioned earlier, the primary user concern relates to the government using the data for purposes other than contact tracing. Therefore, these architectures aim to hide the user identities and generate anonymous IDs for the devices, thereby preventing the ability of the server to link IDs to user information. The decentralised architecture delegates data management to users' smartphones, making the solution more robust against a single point of failure/attack, such as the central server. However, the decentralised architecture still requires a minimally functioning central server. Therefore, it will be vulnerable to a much lower number of  server-based attacks. In decentralised architectures, anonymous IDs are uploaded to the server, which are then potentially accessible by  other smartphones for matching. Thus an honest-but-curious server will not be able to learn any PII, link the anonymous IDs or build social graphs unless it has access to some side-channel information. In case of a data breach, there will be no impact as the attackers  only have access to the seeds/tokens of infected users, which are already public. 
A malicious user, on the other hand, can still cause false positives by relaying the chirps and launch Denial of Service (DoS) attacks by broadcasting fake but correctly formatted advertisements.

The hybrid architecture carries out the contact risk analysis and notification processes at the server. This prevents any re-identification/de-anonymisation attacks, as discussed in Section \ref{sec:Attacks}. In addition, the hybrid architecture provides additional mechanisms to hide user identities from the server while enabling centralised matching of contacts. Similar to the decentralised architectures, it  proposes the generation of ephemeral IDs at the devices. The rationale is that devices keep full control over their secret identifiers, making them less susceptible to breaches at the server.

\section{Proximity Estimation}
\label{sec:proximity}

The chance of becoming infected with COVID-19 increases with prolonged and close contact with an infected person. Hence, estimations of distance (a.k.a proximity) and the duration of contact are important pieces of information used to assess the potential spread of infection.  Contact tracing apps aim to record such encounters.  A smartphone-based automated contact tracing system mainly utilises two (sensor) technologies: GPS and Bluetooth, for proximity estimation\footnote{We note that WiFi can also be used for limited proximity estimation \cite{WiFi}. However, it requires the necessary infrastructure support and setup.}. 
GPS used in navigation systems can provide reasonably accurate location information within the margin of error, especially when used outdoors. However, the storage of absolute location information comes with privacy costs if transferred to the server. Moreover, GPS is not suitable for proximity estimation in COVID apps for several reasons: i) GPS performs poorly in heavily built-up outdoor spaces such as the CBD; ii) GPS does not generally work indoors and, when it works, provides very low accuracy and iii) it consumes power rapidly and can drain the mobile battery quickly if used for a prolonged period.

The Bluetooth interface present in most modern mobile phones is able to capture the Received Signal Strength Indicator (RSSI) values, assisting in proximity estimation \cite{BTProximity}. The wireless signal emitted from a transmitter decays/attenuates as it travels through the air. Equation \ref{eq:1} shows this behaviour where the average $RSSI$ (in dBm) at a distance $d$ from the transmitter is given by
\begin{equation}
\label{eq:1}
    RSSI (dBm) = RSSI_{(d_o)}(dBm) -10 n \mathrm log_{10}(d/d_o)
\end{equation}
where $RSSI_{(d_o)}$ is the average RSSI (in dBm) at the reference distance of 1m, and $n$ is the path loss exponent \cite{rappaport}.
The receiver can approximately estimate how far away the transmitter is by recording the RSSI values. However, there are several issues with proximity estimation solely based on RSSI values. The wireless signal can be influenced by several factors other than distance. The objects in the operating environment\footnote{Some of the environmental factors are captured by variation in the path loss exponent $n$.} such as furniture, walls, people, etc. in the path between a sender and a receiver impact the signal attenuation. Other wireless signals using the same frequency may also cause interference and further attenuate the signal. Besides, different mobile phones transmit Bluetooth signals with different power levels, affecting the distance estimation \cite{rssi}. Finally,  transmission patterns from the same phone vary based on the phone's orientation (specifically its antenna) and the presence or absence of a phone case. These effects can be somewhat mitigated by calibrating the RSSI values/signal attenuation based on known transmit powers.

However, we should be clear that any mitigation strategy aimed at countering the above-mentioned negative impacts on proximity accuracy can only have a limited impact in real-world channel conditions. We must bear in mind that Eq.~\ref{eq:1} represents a simple shadowing model with two free parameters - the path loss exponent and the variance of the noise signal. These unknowns can lead to important uncertainty in the proximity determination. Whilst it is true that a formal location-error analysis on a log-normal shadowing channel can be achieved even in the presence of a-priori unknown values for the path loss exponent and noise variance \cite{malaney}, we do need to understand the limitations of such information-theoretic constructs. Formal analyses of Eq.~\ref{eq:1} leads to Cramer-Rao Bounds on the location accuracy of order a few meters, but  due to the implicit model dependence the accuracy in practice can be substantially larger. Claims of ``guaranteed'' accuracy  of order 1m by any current app should therefore be considered with some scepticism. Along these same lines, we should note there could be other channel models that better describe how signal strength varies with distance in real-world scenarios. It could well be the case that models with more than the two free parameters are required. Further, it is very likely that the usefulness of any such model is dependent on both time and geographical location. However, as we discuss later when we consider future research directions, it should be possible in future tracing apps, built on new hardware platforms (delivering nanosecond timing accuracy), to truly  deliver 1m accuracy with a very high degree of confidence.

To summarise, with the techniques used by current apps for proximity estimation, there would still be many false positives and false negatives. The proximity estimate may indicate close contact, whereas the actual contact is far off or erroneously indicates that it is far off when it is nearby. Similarly, a close contact as perceived by distance estimation does not always translate into an exposed case as there may be a wall/obstruction between the two individuals (e.g., two adjacent apartments), or the contact has occurred in open space where chances of infection are lower. However, getting false positives is not as disastrous, as they only result in additional tests for these false cases. False negatives are a more significant issue as these are considered a missed opportunity to register contact with a positive case. However, we believe that the data from these apps, in conjunction with other contextual information obtained during the interview, would help health professionals to make better decisions.

\section{Attacks}
\label{sec:Attacks}
In this section, we will cover some of the possible attacks that can be launched against different app architectures.

\subsection{Replay/Relay Attack}
\label{sec:replay}
For these attacks, the goal of an adversary is to force the users to store misleading contact information, resulting in false positives. This is achieved by forwarding any message received from honest users at the same or a different location. The adversary requires minimal resources to launch this attack but may use directional antennas to extend the area of its influence further.
A relay/replay attack is the simplest of the attacks that can be launched against users of a tracing app. An adversary can capture the advertised message by a user and immediately relay the captured message at the same location, extending the range of the message, or replay it at another location later on. Note that we classify an attack in this category if the replayed/relayed message has a valid ID (the TempID or chirp); otherwise, it is categorised as a DoS attack (discussed later in Section \ref{sec:dos}).

As the TempID has a short expiry time in centralised systems (Bluetrace recommended 15 min), 
the replay attack can be launched before the expiry of the advertised TempID.
If any person who has received this replay message tests positive, the originator will be identified as a close contact of the affected person (false positive) and may be asked to get tested. A more focused attack is also possible if the replay attack is executed near an epidemic testing clinic or a treatment ward/hospital. Individuals already diagnosed with COVID-19 therefore register the replayed messages as a close contact.


The decentralised version has marked differences in behaviour when viewed with the lens of replay/relay attacks. The chirp generation mechanism, as discussed in Section \ref{De-Chirps},  involves using a seed that is valid for 1 hour. The current timestamp randomises the chirps with 1-minute precision. Finally, the receiver records the time at which each chirp is received. During the tracing process, described in Section \ref{De-tracing}, the app validates the received timestamp of each stored chirp with the time of  creation of each reconstructed chirp, only accepting the received chirp as valid if these two times approximately match. This mechanism provides safeguards against the replay attack. Theoretically, it can still be launched within 1 minute of the chirp message's expiry time. Relay attacks can still be effective as these are not delayed, resulting in valid chirps.

With hybrid architectures, it is still possible to launch relay attacks, as symmetric information would still exist in the PET tables, maintained by two hosts with a malicious relay. However, the replay attacks are not possible as only one of the users would receive the replayed EphID, and the calculated PETs would only exist for the receiver of the replayed message. If that receiver tests positive, the uploaded replayed PETs would not match with any other PET.

Another difference between the centralised and the decentralised architectures w.r.t. the replay attack is the scope of potential targets. In the centralised version, the victim is the originator \emph{(a single person)} of the message being replayed while in the decentralised version, victims are the \emph{multiple} recipients of the replayed message. If the originator tests positive, he/she will upload the seeds to the server (see Section \ref{De-upload}). The recipients of the replay messages will identify themselves as close contacts with the originator by comparing the originator's uploaded encounter chirps. On the other hand, in the centralised version, if any person who has received the replay/relay message tests positive, the originator is falsely identified as a close contact. On the other hand, the relay attack has the same purview in all architectures, affecting both the originator and the recipient of the relayed encounter message.

\subsection{Wireless device tracking}
\label{subsec:device-tracking}
The attacker's goal in this type of attack is to track the device by the BLE information broadcast by the COVID-19 tracing apps. 
Consider a shopping mall that wants to track the general movement pattern of its customers. It can deploy BLE nodes, like Apple's iBeacons, strategically throughout the entire shopping centre, passively listening for advertisements from tracing apps. These nodes can send the captured BLE messages to a central tracking server for further processing. The tracking server can now use simple triangulation~\cite{triangulation} and timestamps to estimate the location of each device. This enables tracking, even recording how much time each customer (device) spends in each store.

For apps that use the centralised architecture, TempIDs and phone model information can be used to to uniquely identify a device. Since TempIDs are changed after a short time (typically 10-15min), tracking a device beyond the point where the device starts advertising a new TempID would require extra intelligence to link the two TempIDs (also see Section \ref{CO Attack})  to the same device, advertising the same phone model. In the decentralised architecture, chirps with a 1-minute lifetime provide limited opportunity for tracking. The tracking server can still enumerate the total number of users in the area, however it is difficult to track the movement of a device without a phone model. The tracking, in this case, would be applicable to limited scenarios e.g., a few customers in a shop or if user’s device is stationery.
Hybrid architectures behave like the centralised architecture as the devices advertise EphID with a lifetime of 15 minutes, making it possible to track a device based on EphIDs.

\subsection{Location confirmation}
\label{sec:locationConf}
In this attack, the attackers' goal is to discover the presence of a user in a known location/environment, such as a neighbourhood. The BLE advertisements and information contained in the exchange of encounter messages in the centralised architecture can be used to confirm a user's location. For example, assume that Alice is the only one in her family who owns an iPhone 9, and this is known to an adversary, Eve. Eve can confirm whether Alice is at home by listening to the encounter messages that include Alice's phone model information. One simple way to mitigate this issue is to include the mobile model number in the registration process. This way, the server still has the required model number information for proximity calculations (see Section \ref{sec:proximity}), and the phone model number can now be excluded from the encounter messages.  A location confirmation attack is not possible in decentralised or hybrid architectures due to the use of ephemeral chirps/IDs and the suppression of user/device linking information.

\subsection{Enumeration Attack}
\label{sec:Enum}
The primary goal of this attack is to count the number of users who have tested positive. Enumeration refers to any user's ability to estimate the number of users infected with COVID-19, who have volunteered to upload their contact tracing data to the server. Note that enumeration does not include the server's ability to count the number of positive cases. In the centralised architecture, the information regarding positive cases and their close contacts remains within the server, therefore preventing users from enumerating.

\begin{figure}[htb]
    \centering
    \includegraphics[width=0.6\textwidth]{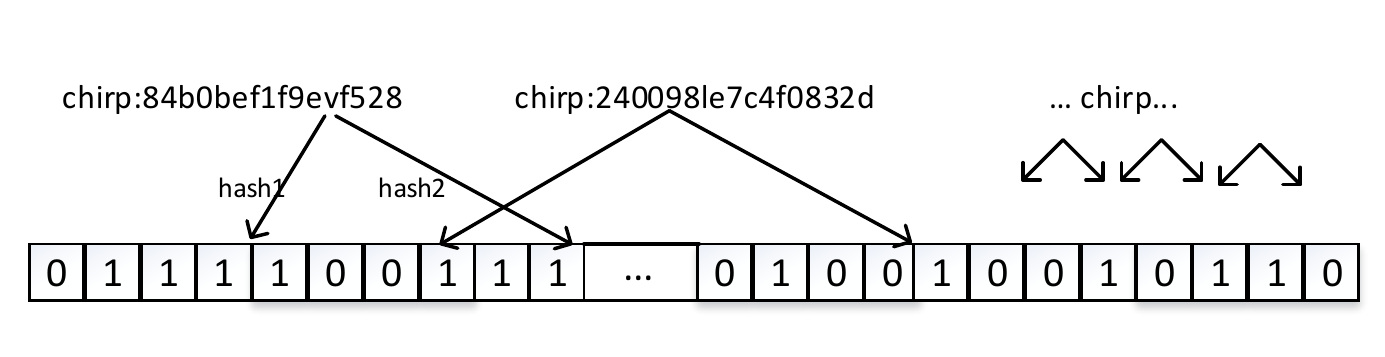}
    \caption{Encoding chirps into a Bloom Filter.}
    \label{fig:bloomfilter}
\end{figure}

In the decentralised architecture, each positive case uploads all of their seed from the last 21 days (21 days x 24 seeds per day = 504 seeds). All app users can download the list of all seeds from the server and can estimate the number of positive cases. One option to conceal this information is to calculate all the chirps at the server and store these in a bloom filter (\cite{Boston} and Section \ref{sec:dp3t-2}). This bloom filter (see Figure \ref{fig:bloomfilter}) is then retrieved by the app to check for matches with their contact chirps, without revealing other details. The enumeration attack can also be mitigated, in the decentralised architecture, if the infected user is provided with the capability to redact some contact information while uploading their contacts. Enumeration attacks are not possible in the hybrid architecture as the server conceals the list of infected user IDs from other users.

\subsection{Denial of Service}
\label{sec:dos}
The goal of this attack is to consume the resources (battery, bandwidth, processing, etc.) available in the system (user mobile, server). In this regard, we discuss the issue of an adversary injecting bogus encounter messages/chirps into the contact tracing environment. This is done with the following, potentially malafide  intentions:

\begin{itemize}
    \item Consume mobile device storage and battery (all three architectures)
    \item Cause an upload of these bogus messages to the server once a user tests positive (centralised and hybrid only)
    \item Increase processing time at the server (centralised and hybrid only)
    \item Increase processing time at the mobile device (more profound in the decentralised architecture as all chirps (including the bogus ones) need to be compared with the reconstructed chirps)
\end{itemize}

\begin{figure*}[htb]
    \centering
    \includegraphics[width=0.9\textwidth]{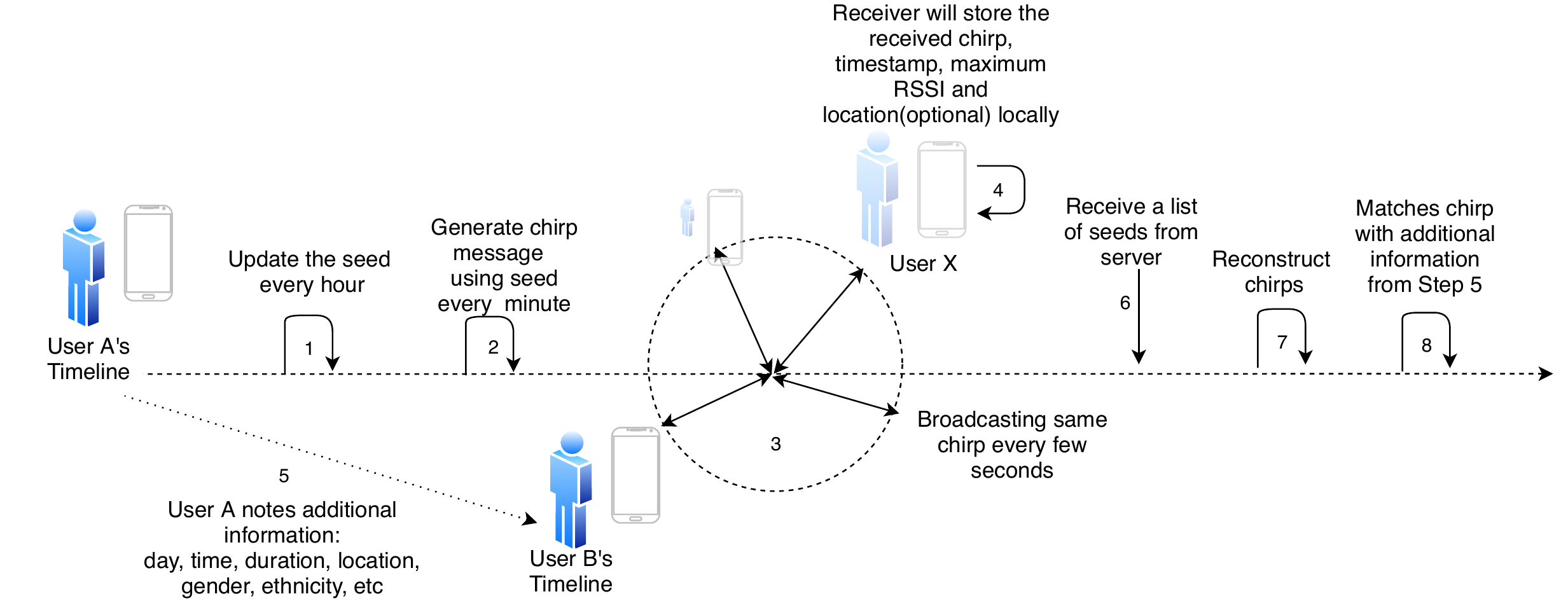}
    \caption{Linkage attack for decentralised architecture.}
    \label{fig:deanonimizing}
\end{figure*}

Note that in the centralised version, the server will process the bogus encounter messages, but will discard these after the server completes a validity check. On the other hand, in the decentralised version, there is no way to check the validity of the received chirp if it is correctly formatted.

\subsection{De-anonymising the users/Linkage attack}
\label{sec:linkage}
In this attack, a user aims to de-anonymise another user's identity by correlating the anonymous broadcast data with information gathered through side-channels. This can be achieved by linking the anonymous ID with the user's identity in what is known as a linkage attack. Most contact tracing apps have been designed with data and user privacy in mind. However, in the decentralised architecture, it is still possible to identify users once they test positive to the virus. Figure~\ref{fig:deanonimizing} presents the steps required to launch the attack. User A uses a decentralised app that records the details of his encounters with other persons (day/time/duration/location/gender, etc.) (step 5). If this user receives an alert (step 6), he/she can easily identify the infected user by comparing the reconstructed chirps (step 7). This can be achieved by looking at the time stamp (and duration) of the chirps and comparing his/her collected records (step 8). Some malicious record keeping can also be done automatically by a modified app that collects location information using GPS/WiFi etc.

For the centralised architecture, it is possible to de-anonymise close contacts, but it is hard to de-anonymise a positive case since an app user is not provided with a list of TempIDs for comparison. A positive case can still be identified if a user who is in isolation and has only met one person receives a close contact notification. TempIDs can easily be associated with a user by referring to the advertised mobile model number. The duration of contact and an isolated encounter will increase the chances of linking TempIDs with a particular user. Similarly, a Sybil attack~\cite{douceur2002sybil} can also be launched whereby an attacker can deploy multiple devices and only use a single device for a short time. If the user receives a notification from the server on one of his/her devices, the user can narrow the linkage attack to a short time window when that device was active.

An attacker can launch another kind of linkage attack, called a Paparazzi attack \cite{V2020} \cite{V2020a} in decentralised apps using passive BLE devices. When a user tests positive, the server receives the seeds, which are, in turn, sent to the users, including the attacker. The attacker reconstructs chirps and combines this data with that obtained from the passive BLE devices. It can then track the positive case throughout the contagion period. Similar to the Paparazzi attack, attackers can trace infected users by deploying a large number of passive BLE devices while colluding with the server. This is referred to as an Orwell attack \cite{ABIV2020}.

Hybrid protocols are generally considered un-linkable as these do not share the PETs from infected users with other users, and they perform risk analysis and notification in a similar manner to the centralised version.

\subsection{Abuse of app}
\label{sec:abuseApp}
The goal of this type of attack is to mislead the tracing app with information produced through incorrect usage of the app. All tracing apps, whether based on centralised, decentralised, or hybrid architectures, are prone to user abuse. A recent Google Maps experiment is referred to, in which a user carted 99 mobile phones down an empty road, and Google Maps responded by showing high congestion \cite{GmapsHack}. Similarly, a tracing app cannot recognise if a phone is being carried by its owner, someone else, or tied to a pet running in the park \cite{oxymoron}.

New measures could be introduced to circumvent these activities, such as using other existing sensors on the phone to record activity or perform gait recognition. However, from a privacy perspective, capturing greater quantities of contextual or personal data may result in privacy guarantees being lowered.

\subsection{Carryover attack}
\label{CO Attack}
This attack is supplementary to the device tracking attack discussed in \ref{subsec:device-tracking}. An attacker aims to continue the device tracking period beyond the anonymous ID expiration time. Most devices randomise their Bluetooth MAC addresses to avoid device tracking. The temporary identifiers (TempIDs and chirp) also get changed after a short time. An address carryover attack \cite{becker2019tracking} is possible when the changeover time of a Bluetooth MAC address and the temporary identifier are not synchronised. For example, let us assume that the TempID gets changed every 15 min, while the Bluetooth MAC changes every 10 min. A listener can easily link the multiple Bluetooth MAC addresses advertised within the lifetime of the same TempID. Conversely, a TempID change can be linked to the same Bluetooth address. Full synchronisation of temporary ephemeral identifiers and the Bluetooth random MAC address change are proposed in some of the apps. The changeover synchronisation, however, requires support from the OS Bluetooth module.

Note that even when the changeover of multiple identifiers is fully synchronised, a careful listener can still link different Bluetooth addresses (or temporary IDs) for wireless tracking by analysing the disappearance of an identifier and its immediate replacement by a new one. However, in our vulnerability analysis of apps (in Section \ref{sec:apps}), we do not consider this case.

\subsection{Disclosure of social graph }
\label{sec:socialGraph}
A social graph illustrates the interaction between individuals by representing the individuals as nodes, and connection between the nodes as edges indicating that these users may have been in close proximity. An attacker can build the social graph by mining the available data to infer the contact profiles of users. Disclosure of the social graph or even a part of it is undesirable, however some countries such as India have done so despite concerns from civil society. Both centralised and decentralised systems are vulnerable to the disclosure of (partial) social graphs.

In a decentralised system, if a malicious server wants to know whether an infected user A and target user B were in contact, it sends the seeds uploaded by A along with some other fake seeds to B. Now, by using a side-channel, if it observes that B has received an alert or gone into isolation, then it can say that A and B had been in contact. In the centralised architecture, positive cases can choose to upload their contacts to the server. A malicious server can construct part of the social graph (positive cases and their close contacts) because it knows the mapping between the TempIDs and individuals. Construction of a complete social graph is impossible unless the server is powerful enough to control the communication network and later perform huge data mining tasks. In any of the discussed architectures, it is to be noted that a malicious server cannot disclose the interaction between individuals who neither tested positive nor have been in contact with other positive patients. Theoretically, an attacker could steal several phones and possibly correlate symmetric encounters across users.

To hide the social graph construction by the server, apps from the centralised and hybrid categories have introduced additional measures (such as random, independent uploads of contact identifiers/EphIDs, maintaining separate upload and query identifiers, using zero knowledge proofs \cite{ZKP} to store anonymized social graph at the server~\cite{792} etc.). Some of these measures will be discussed in Section \ref{sec:apps} where we explore different protocols and apps in detail.

\section{Analysis of Specific Apps and Protocols}
 \label{sec:apps}
\begin{figure*}[htb]
    \centering
    \includegraphics[width=0.65\textwidth]{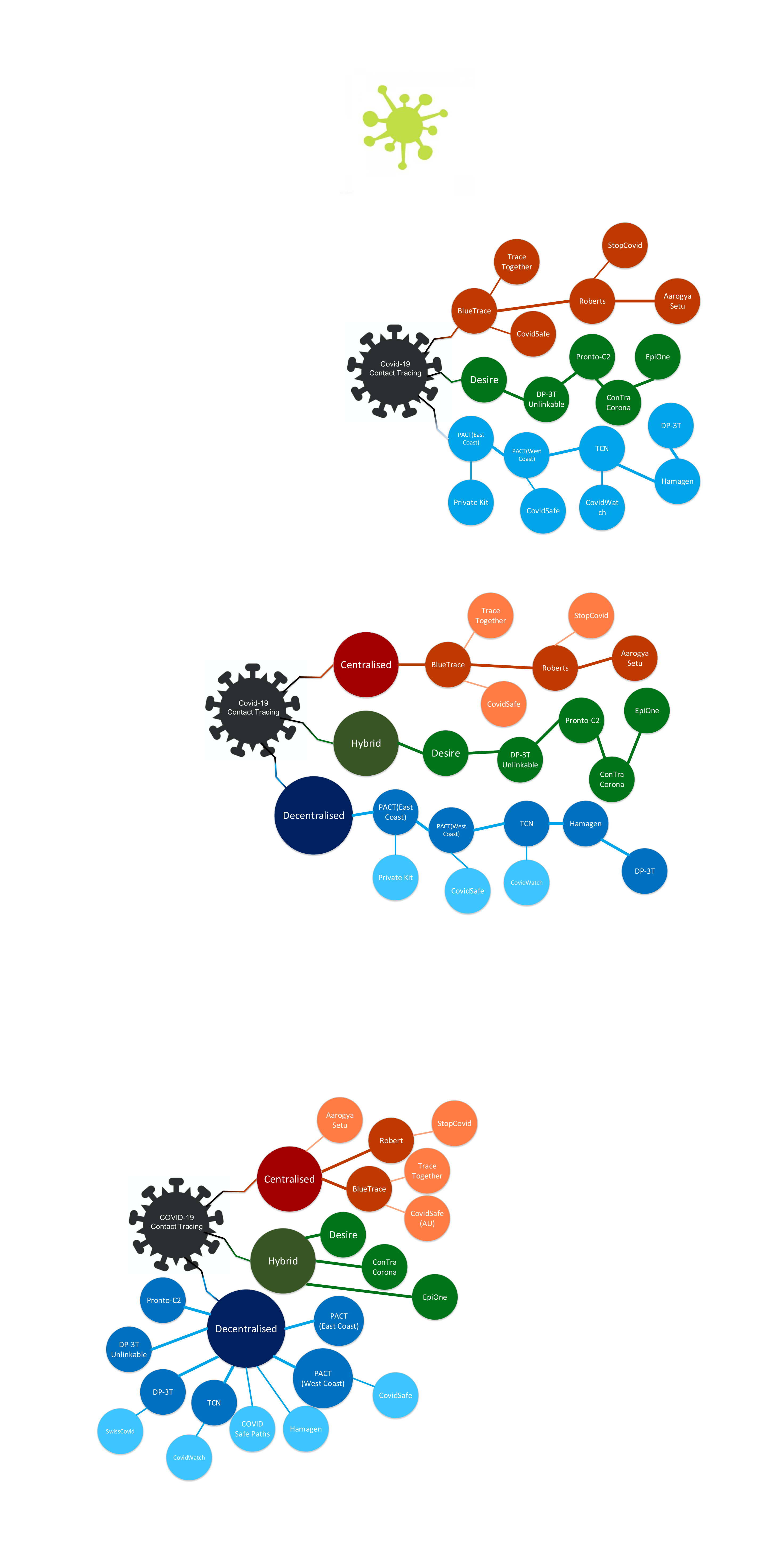}
    \caption{Summary of apps and protocols.}
    \label{fig:covid-summary}
\end{figure*}

Above, we discussed the salient features, capabilities, and exposure to different kinds of attacks for the three broad architectures, which have been proposed for contact tracing applications. In this section, we delve into the instantiation of these architectures and introduce various tracing apps and protocols that are being proposed, developed, and deployed in many countries. We divide this discussion into three sections, each covering a different system architecture.  To keep the discussions concise, we only focus on the salient features of each app/protocol and highlight any additional measures introduced on top of the corresponding base architecture.  Figure \ref{fig:covid-summary} illustrates the protocols and the apps discussed in this section. Additionally, we provide an attack matrix (Table \ref{table:Attacks}) that highlights possible attacks on each app/protocol that is part of our discussion.

\subsection{Apps/Protocols based on centralised architecture}
In Section \ref{sec:Arch}, we explained the centralised architecture that was based on the Bluetrace \cite{bluetraceWhitepaper} protocol. ROBERT \cite{robert} is a similar protocol that relies on a centralised architecture. We will provide an overview of the TraceTogether (Singapore) \cite{opentraceGit} and CovidSafe (Australia) \cite{covidsafeGit} apps which are based on the Bluetrace protocol, and the StopCovid (France) app \cite{stopcovid} which implements the ROBERT protocol. The Aarogya Setu (India) is another instantiation of the centralised architecture and uses both Bluetooth and GPS.

\subsubsection{TraceTogether}
\label{sec:traceTogether}
The TraceTogether app launched by the Singaporean government in March 2020, was among the first contract tracing apps deployed to the general public. The source code of the reference implementation, OpenTrace, has been released in the public domain \cite{opentraceGit}. We only detail one additional BlueTrace protocol feature that has been implemented in the TraceTogether app but which was omitted in the centralised architecture description presented in Section \ref{sec:Arch}. The server issues forward-dated TempIDs to each device instead of a single TempID. This is to ensure that each device has a supply of valid TempIDs even when the Internet connection is unstable.


\subsubsection{CovidSafe (AU)}
\label{sec:covidSafe}
CovidSafe was released by the Australian government on 26th April 2020, with the source code for both Android and iOS clients released on 8th May 2020 \cite{covidSafe-Source}. Covid-Safe follows the Bluetrace protocol and exhibits many similar characteristics to the TraceTogether app, including vulnerability to the different types of attacks listed in Table \ref{table:Attacks}.

The two apps differ in the lifeTime of TempIDs. TraceTogether
uses a value of 15 minutes as recommended in the BlueTrace protocol specifications, while CovidSafe employs 2 hours. This makes CovidSafe more vulnerable to replay attacks. The relative advantage of CovidSafe over TraceTogether is that the devices do not have to download TempIDs frequently. Another difference between the two apps is in the backend infrastructure. While TraceTogether uses Google Cloud to provide the backend services, CovidSafe makes use of Amazon AWS servers that are located within Australia.

\subsubsection{StopCovid - ROBERT}
\label{sec:stopCovidROBERT}

ROBust and privacy-presERving proximity Tracing protocol (ROBERT) is a centralised tracing app protocol, that is jointly developed by researchers at INRIA (France) and Fraunhofer (Germany)~\cite{robert}. The StopCovid app, which is under development in France at the time of writing and is expected to be available in June 2020,  will use the ROBERT protocol. The source code for the StopCovid app is already available \cite{stopCovid-Source}.

The main difference between the BlueTrace and ROBERT protocols is the type of user data that is stored on the server. While the former stores PII, the latter only stores anonymous identifiers referred to as EphIDs, thus providing a level of privacy. 
The protocols also differ in the notification process, i.e., how the at-risk users are notified. ROBERT requires all users to frequently check their used EphIDs with the server to determine if they are flagged as at-risk. In contrast, with BlueTrace, health authorities could proactively notify at-risk users. The notification process is possible because BlueTrace can map the TempIDs to the collected personal information.

In ROBERT, a positively identified user uploads the EphIDs in a staggered and random order. 
This is in contrast with the BlueTrace protocol, where all contacts are uploaded in one go. This is done to break the link of contacts with the same person and to prevent the server from conducting social graph analysis. However, analysing the network traffic exchanged by a device may potentially allow an adversary to link the reports together.

\subsubsection{Aarogya Setu}
\label{sec:aarogyaSetu}
Aarogya Setu is an app deployed in India based on the centralised architecture. 
The code for their Android app was released (iOS and server code not yet available) on 26th May 2020 \cite{setu-Source}.
In addition to collecting PII and contact data, this app also gathers location data (GPS coordinates) and self-assessment data (responses provided by an individual to the self-assessment test)~\cite{setu2}.
It performs data analytics on the gathered information to indicate how many positive cases are within 500m to 10km of a user's current location. Users can upload the trace data if they are COVID-19 positive, or they fail the self-assessment test. 
In contrast to other apps, the government of India has made it mandatory for all government employees and those living in disease containment zones to install the Aarogya Setu app.

\subsection{Apps/Protocols based on decentralised architecture}
In this section, we first briefly discuss the Apple/Google alliance for providing APIs and OS-level support for privacy-preserving contact tracing. Next, we provide a short review of the Private Automated Contact Tracing (PACT) protocol by MIT \cite{pact-ec}, which was used as the basis for describing the decentralised system architecture in Section  \ref{sec:dec}. Another protocol sharing the same name, PACT (Privacy-sensitive protocols And mechanisms for mobile Contact Tracing), developed by a team from the University of Washington \cite{chan2020pact}, is also discussed. To distinguish the two PACT protocols, we use the nomenclature:  “PACT West-coast” (by UoW) and  “PACT East-coast” (by MIT). Other contact tracing protocols including DP-3T \cite{DP-3T}, TCN \cite{TCN}, Hamagen \cite{Hamagen}, COVID Safe Paths \cite{pksp} and Pronto-C2 \cite{ABIV2020} are also covered in this section.

\subsubsection{Apple/Google Exposure Notification APIs}
Apple and Google \cite{Apple, Google} have joined hands to support privacy-preserving contact tracing by developing an exposure notification system. Their proposed tracing mechanism matches with the decentralised architecture described earlier in Section \ref{sec:dec}. Their system was planned to be rolled out in two phases. During the first phase, 
on 20th May 2020, APIs were released to support apps developed by health authorities, intended to work seamlessly on iOS and Android devices. This is designed to help manage issues related to BLE scanning and advertisements faced by current apps \cite{bluetraceWhitepaper}. Apps based on recently released APIs by Apple and Google can use the proposed Associated Encrypted Metadata (AEM) during the exposure notification service. AEM is a privacy-preserving encrypted metadata that includes transmit power to aid in collecting more accurate proximity estimation results.

In the second phase, Apple/Google plan to incorporate OS-level support to help broader adaptation by eliminating the need for an app to perform contact tracing.
An app called SwissCoviD, discussed later in Section \ref{sec:swissCovid}, was released on 25th May 2020 for pilot testing by the DP-3T team \cite{DP-3T}. SwissCoviD is based on the Apple/Google APIs. Similarly, an open-source app called Corona-Warn-App \cite{CoronaWarn} has been released on 16th June, 2020 in Germany that is based on these APIs. Another project named Aurora \cite{aurora} is under development by the PathCheck team to utilise the APIs developed by Apple/Google. Moreover, many existing apps, including some from the centralised category, have already started exploring ways to migrate their existing code base to the APIs released by Apple/Google.

\subsubsection{PACT (East-coast)}
\label{sec:PACT-EAST}
The PACT (East-coast) protocol design was used as the basis to explain the decentralised architecture, including installation, encounter exchange, and the tracing process in Section \ref{sec:dec}. A research collaboration led by MIT~\cite{pact-mit} developed this protocol.

In addition to locally stored seed data (used to generate chirps) and chirp data (including received time and its signal strength), the PACT (East-coast) also allows the user to optionally store extra metadata, such as location information, in its local log file while receiving a chirp. The location data can help determine the place of contact, for example, a restaurant or a park. This optional metadata can help reduce false positives and increase system accuracy by involving more contextual information.




%

\subsubsection{CovidSafe - PACT (West-coast) }
\label{sec:PACT-WEST}
Privacy-sensitive protocols And mechanisms for mobile Contact Tracing - PACT (West-coast) \cite{chan2020pact} is a tracing protocol proposed by researchers from the University of Washington.
The core mobile tracing function uses a very similar process, compared with the baseline decentralised systems used in the PACT East-coast app, in which all personal data is locally stored (and encrypted) on the phone, devices broadcast pseudorandom IDs, and it is voluntary for a user to publish/upload the data. PACT (West-coast) adopts a different key-based generation mechanism to generate the pseudorandom ID used for broadcasting. A 128-bit key is initially fed into the pseudorandom generator, and a 256-bit length output is generated (take SHA-256 as an example). The half-length of that result (128-bit) is used as the temporary pseudorandom ID  within a specified period $[t_0+dt \cdot i, t_0+dt \cdot (i+1)]$ and the other half is used as the input for the next pseudorandom generator. This key-chain design saves storage, storing fewer seeds than the PACT (East-coast) app. Afterwards, those pseudorandom IDs are broadcast and collected, similarly to the PACT (East-coast).

Like all decentralised protocols, PACT (West-coast) is also susceptible to the linkage, and enumerations attacks (discussed in Section \ref{sec:Attacks}), as the seeds from infected users are uploaded to the server and later on shared with all users.
CovidSafe (UoW) app~\cite{uow-CovidSafe}, for which the beta version was released in May 2020, is based on the PACT (West-coast) protocol. 

\subsubsection{SwissCoviD - DP-3T}
\label{sec:swissCovid}
Decentralised Privacy-Preserving Proximity Tracing (DP-3T) \cite{DP-3T} is a protocol specification based on the decentralised architecture proposed by a consortium of universities and organisations from Europe led by EPFL, Switzerland.  A pilot app named SwissCovid \cite{swissCovid}  was released on 25th May, 2020\footnote{This is the first official release that is based on APIs developed by Apple/Google}. There are two versions covered in the specifications, we explain the `low cost' version here, while the other `un-linkable' version is covered in the next section.

The protocol is very similar in functionality to the base decentralised architecture discussed earlier (Section \ref{sec:dec}). A daily key is generated by each device, using the hash-chain from the previous daily key, and then used to generate the EphIDs. Messages are exchanged with other devices which come in contact over Bluetooth containing these EphIDs with expiry information and coarse time (date). The exchanged data remains in local storage until a user tests positive. Health officials determine the contagious window, i.e., at what time a positive case is contagious and might infect others. Identified users only upload their daily keys starting from this date. Once data is uploaded, the infected user changes their random key for generating future daily keys to prevent future tracking/identification. Other users download the daily keys uploaded by the infected user and compare their stored EphIDs with EphIDs reconstructed from the infected person's daily keys. The risk analysis process is thus performed locally on individual devices. The app notifies the user if they are found to be in close contact and asks for permission to upload their daily keys. 

\subsubsection{DP-3T Unlinkable}
\label{sec:dp3t-2}
The DP-3T specification document \cite{DP-3T} also has a second version called an `Un-linkable' design. This design is in response to the claims that the decentralised design is subject to Linkage and Enumeration attacks, discussed in Section \ref{sec:Attacks},  as the daily keys for infected users are made available to all other devices.

The primary change in this design is that the daily keys uploaded by infected users are converted into their corresponding EphIDs by the server. The server hashes these values in a Cuckoo filter \cite{cuckoo} before advertising these to the other users. The users can still check whether their received EphIDs that are cryptographically hashed with the received time, and whether these match any entry in the Cuckoo filter or not. The user cannot access other information, including how many or which other EphIDs have been encoded in the Cuckoo filter.

The processing time at the server increases compared with the low-cost version of DP-3T. Also, the Cuckoo filter needs to be carefully designed to minimise the chances of false positives, while false negatives are not possible with this filter. 
This design also enables users to selectively upload encounters (by suppression/redaction of some encounters) in the upload phase.

Another proposed feature is to use k-out-of-n secret sharing to minimise the chance of EphID collection during short periods of contact. A contact has to collect at least k advertisements to reconstruct an EphID successfully. This
restricts the effective EphIDs to contacts that are made for a sufficient duration.

\subsubsection{CovidWatch - TCN}
\label{sec:covidWatchTCN}
Researchers from Stanford University and Waterloo University have been developing Covid-Watch \cite{covidWatch}. At the time of writing, this app is still in a pilot phase. The source code is publicly available~\cite{CovidWatchSource} and it follows the TCN (Temporary Contact Number) Coalition protocol~\cite{TCN}.

The TCN protocol generates the keychain, such that each key (seed) derived from the master key generates one unique temporary contact number (chirp). Like other decentralised protocols, TCN only uploads the compact seed data (the Master key, the expiry time, etc.) to the server rather than the entire list of TCNs. All seeds that belong to a report can be proven/verified by the server as they are generated and bound to the same master key. Malicious entities can potentially launch a linkage attack (see Section~\ref{sec:linkage}) to find out the linkable TCNs by observing multiple TCNs from the reports that use the same master key. Therefore, frequent master key rotation can be performed to make TCNs un-linkable from different reports. However, this increases the number of keys that need to be maintained, raising the issue of scalability. An app thus needs to consider the trade-off between scalability and linkability while selecting the master key rotation period.


\subsubsection{Pronto-C2}
\label{sec:prontoC2}
Researchers from the University of Salerno, Italy proposed the Pronto-C2 contact tracing system \cite{ABIV2020}. It is a decentralised app that lets the devices communicate anonymously with each other while hiding these communications from the central server, averting mass surveillance.

At the heart of the protocol are two cryptographic tools: the Diffie Hellman (DH) Key exchange \cite{DH} discussed in Section \ref{sec:hybrid} and blind signatures \cite{C83}. A  secret $sk_A\in Z_p$ and the ephemeral ID  $eph_A = g^{sk_A}$ are generated by Alice's device and stored in the server. The address $addr_A$ is noted by the device where its generated ephemeral ID is stored on the server (this storage can also be realised using a  blockchain). 
The device broadcasts this address and receives addresses from other devices in its vicinity. This is contrary to decentralised designs that share ephemeral IDs. Alice stores the  tuple $(eph_{A,i}, sk_{A,i}, addr_A, t)$ when it receives the $eph_B$ from Bob. Here $sk_{A,i}$ is the secret key from the previous update $i$, and $t$ contains auxiliary information like BLE signal strength, time, etc.

If Alice tests positive, she fetches the ephemeral ID of each contact from her contact list. For Bob, a contact, she receives $eph_B$ using the $addr_B$. Alice  calculates $K'= eph_{B}^{sk_A}$ which is the DH key between herself and Bob, and a key $K = H(K'||eph_A||eph_B)$. She then sends the (blinded) information of $K$ to an authentication server, which appends a blind signature to $K$. The signature prevents DoS attacks and ensures that the authorisation server cannot perform social graph analysis. 

A user $Y$ who wishes to test their risk can download the ephemeral IDs from their contact list in address $addr_X$. They then compute the DH key $K'$ between themselves and the ephemeral ID $Eph_X$ at $addr_X$. It computes $K= H(K'|| Eph_X||Eph_Y])$. $Y$ downloads the recently available keys (of infected individuals) from the server and checks if $K$ belongs to this set. If a match is found, then the user is at risk. The devices communicate with the server using anonymous channels like TOR to prevent linking ephemeral keys to users.


\subsubsection{Hamagen}
\label{sec:hamagen}
The Hamagen application is developed by Israel's Ministry of Health. Hamagen is different from many other contact tracing applications in that it does not rely on recording encounters with other phones in the vicinity using Bluetooth. The application cross-checks the GPS history of a mobile phone with the historical geographical data of identified cases from the Ministry of Health. The check is conducted locally on the individual's mobile phone. Each individual's location data does not leave their device, and neither is it sent to a third party. The app will periodically (currently set to hourly) download a file containing an anonymised list of locations that were visited by individuals diagnosed with COVID-19 over the past 14 days. This file is sourced from the Ministry of Health and populated with data showing those people who have undergone epidemiological investigation using the various tools available to the Ministry. The application then cross-references these locations (including timestamps) with the location data stored locally on the individual's device. Should the application discover that there is a possibility that the individual has been at the same place and at the same time as a diagnosed case, a notification is shown on the phone with the details of the location and times where the individual may have been exposed to a positive case. The phone user has the option to review the notification. If the message is perceived to be incorrect, e.g., if the user was not at the noted location at the stated time, the user can indicate that the information is false. If the user confirms their presence at the contact location, they are directed to the Ministry of Health's website for information on what to do next.

The application must be given access to the location history (GPS data) of the phone and the list of cellular base stations and WiFi access points encountered over the past two weeks. This information is stored in SQLite. The app also requires Internet access to periodically download an updated file from the Ministry of Health server. The application source code has been developed using React Native and is open-source on GitHub~\cite{Hamagen}.

\subsubsection{COVID Safe Paths}
\label{sec:COVIDSafePaths}
The PathChecks team \cite{pathCheck} has developed this app and its source code for Andriod and iOS and has made it publicly available \cite{pksp}. The app is similar to the Hamagen in functionality as it also employs logging of GPS location trajectories. A browser-based map tool called ``Safe Places" has also been released that can interact with the Safe Paths app. The diagnosed users can voluntarily share their location trails with the health authorities using the Safe Places map tool. Other users can download the anonymised and aggregated data sets of public locations to check whether they have come in contact with an identified individual, without uploading their path trajectories.


\begin{table*}[ht]
  \footnotesize
  \centering
  \caption[Attacks]{Possible Attacks on Tracing Apps and Protocols \footnotemark}
  \label{table:Attacks}
    \begin{tabular}{ |c|c|c|c|c|c|c|c|c|c|}
      \hline
      Section & Tracing & Replay/ & Wireless & Location & Enumeration & DoS & Linkage  & Carryover & Social\\
      No. & Apps \& Protocols & Relay & tracking & confirmation & & & & & graph
      \\
      \hline
      \ref{sec:traceTogether} & Trace Together &\Checkmark  & \Checkmark & \Checkmark & $\times$ & \Checkmark& \Checkmark& \Checkmark &Easy
      \\
      & (BlueTrace) & & & & & & & &
      \\
      \hline
      \ref{sec:covidSafe} & CovidSafe (AU) &\Checkmark  & \Checkmark & \Checkmark & $\times$ & \Checkmark& \Checkmark& \Checkmark &Easy
      \\
      & (BlueTrace) & & & & & & & &
      \\
      \hline
      \ref{sec:stopCovidROBERT} & StopCovid &\Checkmark  & \Checkmark & $\times$ & $\times$ & \Checkmark& \Checkmark& $\times$ &$\times$
      \\
      & (ROBERT) & & & & & & & &
      \\
      \hline
      \ref{sec:aarogyaSetu} & Aarogya Setu  & \Checkmark & \Checkmark & \Checkmark  & \Checkmark   & \Checkmark & $\bigcirc$ &$\bigcirc$ & Easy
      \\
      \hline
      \hline
      \ref{sec:PACT-EAST} & PACT (East Coast) & Limited Replay & \Checkmark & $\times$ & \Checkmark & \Checkmark & \Checkmark & \Checkmark & Difficult
      \\
       & & \Checkmark Relay  &  &  &  & & &  &
      \\
      \hline
      \ref{sec:PACT-WEST} & CovidSafe (UoW)&Limited Replay & \Checkmark  & $\times$ & \Checkmark &\Checkmark  & \Checkmark  & $\times$ & Difficult
      \\
      & (PACT-West Coast) &\Checkmark Relay & & & &  & & &
      \\
      \hline
     \ref{sec:swissCovid} & SwissCovid - DP-3T &\Checkmark &\Checkmark & $\times$ & \Checkmark &\Checkmark &\Checkmark &\Checkmark &Difficult\\
     & (low cost) & & & & & & & &
      \\
      \hline
    \ref{sec:dp3t-2} & DP-3T &\Checkmark & \Checkmark & $\times$ & $\times$ &\Checkmark & $\times$ &\Checkmark &Difficult\\
      & (unlinkable) & & & & & &  & &
      \\
      \hline
     \ref{sec:covidWatchTCN} & CovidWatch  & \Checkmark & \Checkmark & $\times$  & $\times$  & \Checkmark &\Checkmark & $\times$   &Difficult
      \\
      & (TCN) & & & & & & & &
      \\
      \hline
     \ref{sec:prontoC2} & Pronto-C2  &\Checkmark &\Checkmark  & $\times$ &$\times$  & \Checkmark  &\Checkmark& $\bigcirc$ &$\times$
      \\
      \hline
     \ref{sec:hamagen}  & Hamagen & $\times$ & $\times$ & $\times$ & $\times$  & $\times$ & \Checkmark & $\times$  & $\times$
      \\
      \hline
     \ref{sec:COVIDSafePaths}  & COVID Safe Paths & $\times$ & $\times$ & $\times$ & $\times$  & $\times$ & \Checkmark & $\times$  & $\times$
      \\
      \hline
      \hline
     \ref{sec:desire} & DESIRE  &\Checkmark Relay only & \Checkmark & $\times$ & $\times$ &\Checkmark &$\times$ & $\times$ & Difficult
      \\
      \hline

      \ref{sec:contracor}& ConTra Corona  & \Checkmark  & \Checkmark & $\times$ & $\times$ & \Checkmark & $\times$ & $\times$ & Difficult
      \\
      \hline

     \ref{sec:epione}  & EpiOne  & \Checkmark& \Checkmark &$\times$ &$\times$ & \Checkmark  &\Checkmark &\Checkmark & $\times$
      \\
      \hline
    \end{tabular}
  \end{table*}

\footnotetext{Abuse of the app (Section \ref{sec:abuseApp}) is common for all apps/protocols, hence not shown in Table \ref{table:Attacks}. $\bigcirc$ denotes not enough information is available}
\subsection{Apps/Protocols based on hybrid architecture}
Hybrid protocols (Section \ref{sec:hybrid}) have been proposed to combine features of both centralised and decentralised architectures. We discuss three reference protocols in this section, DESIRE \cite{Desire}, ConTra Corona \cite{contra} and EpiOne \cite{epione}.

\subsubsection{DESIRE}
\label{sec:desire}
The explanation of the hybrid architecture in Section \ref{sec:hybrid} is based on the DESIRE protocol specification \cite{Desire}. The use of cryptographically generated PETs gives users more control while keeping these different from the advertised EphIDs. This avoids the potential harvesting of contact data for social graph analysis. All data stored at the server is encrypted with keys which are stored at the clients' devices. This protects the client data in case the server has a data breach.

The risk analysis and notification are handled by the server (instead of the clients as in the case of decentralised versions), which limits the likelihood of  other users launching Enumeration and Linkage attacks.

\subsubsection{ConTra Corona}
\label{sec:contracor}
ConTra Corona\cite{contra} is a hybrid protocol proposed by German researchers from the FZI Research Center for Information Technology and Karlsruhe Institute of Technology. Contra Corona improves privacy protection by mitigating the linkage attacks that can be launched on decentralised apps. This is achieved by adopting a DDH key-exchange mechanism to verify the data upload process for a person diagnosed with COVID-19. Additionally, Contra Corona proposes strict server separation by employing three different servers; the submission server, the matching server, and the notification server.

The Contra Corona proposal is significantly different from the base hybrid architecture (based on the DESIRE protocol). However, the underlying assumptions are the same. Devices generate their IDs, and the centralised server(s), are responsible for the risk-analysis and notification process.



Each user generates a warning identifier ($wid$) for each day based on their real identifier (e.g., name). For each $wid$, the device computes 96 $sid$'s (regarded as seed identifiers) encrypted with the submission server's public key and $pid$ (pseudorandom identifier encrypted and hashed based on $sid$). $pid$ and $sid$ are uploaded to the submission server. Initially, the submission server collects all users' $(sid,pid)$ pairs. Once the submission server has accumulated  sufficient client pairs, it shuffles these and then sends them to the matching server (this helps to mitigate the enumeration attack). If the matching server receives a $pid$ uploaded by one of the infected users, the matching server looks up the respective $sid$ of all potentially contaminated users and sends them to the notification server. Finally, the notification server decrypts the $sid$ to recover the user's warning identifier ($wid$) and publishes the $wid$ list. All users regularly fetch the $wid$ list from the notification server and compare it with the $wid$s (stored locally) that they have used for the previous 28 days.

Contra Corona protocol utilises n-out-of-k secret sharing of $pid$ (n is taken as 15 and k as 45) and selects a random identifier $m$ (for example, $m$ may be taken as the Bluetooth MAC address). One secret share of $pid$ is broadcasted every minute. Users who received and accumulated 15 such broadcasts sharing secrets of the same user $m$ can reconstruct the $pid$ of that contact event and further store that contact event on the device.


ConTra Corona's privacy enhancement is based on the premise that the servers are non-colluding, and all communication channels are anonymised or authenticated. The submission server is assumed to be the most trusted component as it stores all the matching $(sid,pid)$ pairs of identifiers that are generated by each user. The additional encryption and randomisation are used to prevent disclosure of infected user's status to other users. The protocol also requires the health authority to verify the source integrity of the report (uploaded by the infected person) to decrease the false negatives in the system (and also prevent the DoS attack).

\subsubsection{EpiOne}
\label{sec:epione}
EpiOne \cite{epione} has been proposed by a team of researchers led by the University of California at Berkeley to protect against linkage and social graph analysis attacks discussed in Section \ref{sec:Attacks}. At the time of writing, the source code has not been made available. At the heart of the protocol is a cryptographic technique known as Private Set Intersection (PSI)\cite{PSI, CLR17}.

At a high level, devices generate and share random IDs (tokens) using a seed. Each device maintains sent and received token lists. If a user tests positive, they upload their seeds through the health officials who use the server to construct their sent tokens. A user who wants to check his close contacts and server must follow a private set intersection protocol to check if there is an intersection between received tokens and those maintained at the server. The user is notified once a match is found. The private set intersection protocol guarantees that neither the user nor the server knows the complete set of tokens, thus preventing social graph analysis.

At a more fundamental level, EpiOne consists of two servers, a collection server with the healthcare officials and an untrusted verification server.
Before the start of the protocol, all participants, including the users, healthcare professionals and verification server first agree on the security parameters.
The verification server generates a public/private secret key pair and publishes the corresponding public key. Every user can choose a random seed and derive the tokens using input seed, the day, and the time slot within the day. This is useful because the verification server can, later on, reconstruct the token if it knows the corresponding seed.

When users come into contact with each other, they exchange tokens. The list of tokens sent and received are stored on the mobile device. When a user tests positive, it uploads the encrypted seeds (encrypted with the verification server's public key) to the collection server maintained by health officials. The health officials, in turn, shuffle the encrypted seeds and send them to the verification server that can decrypt the seeds and reconstruct all the tokens.

A user can find out if they were in close proximity to a positive COVID case by computing securely (using Private set intersection protocol) the cardinality of the intersection between the two sets of tokens. The server neither reveals the set of tokens of positive cases (preventing enumeration attack), nor does the user reveal the set of tokens they received from other users (preventing social graph construction).

\begin{table*}[ht]
  \footnotesize
  \centering
  \caption{Comparison of factors affecting device battery usage}
  \label{table:Battery}
    \begin{tabular}{ |c|l|l|l|l| }
      \hline
      Architecture & Encounter exchanges & Downloads from server& Data upload to server & Processing at device
      \\
      & & & (Once for a positive case) &
      \\
      \hline
      Centralised & BLE periodic exchange & Periodic download & Upload all encounter & Minimal processing
      \\
      & of short messages& of TempIDs & messages for the &
      \\
      & (TempIDs) &(once in 15 min) & past 21 days &
      \\
      \hline
      Decentralised &BLE periodic exchange & Download of seeds& Upload seeds used & High processing
      \\
      & of short messages& for positive cases& for the past & device periodically generates
      \\
      & (Chirps) &(once in 24 hours) & 21 days & seeds/chirps
      \\
      \hline
      Hybrid &BLE periodic exchange& - & Upload PETs used for & High processing
      \\
      & of short messages& &the past 21 days & device periodically generates
      \\
      & (EphIDs) & & *Upload PETs for risk & EphIDs and calculates,
      \\
      &  & &  analysis by server & two PETs for each EphID
      \\
      \hline
    \end{tabular}
  \end{table*}

\section{Common user concerns}
\label{sec:commonconcerns}
The COVID-19 tracing applications have seen increased adoption in many  countries; for example,  more than 5 million users  downloaded the CovidSafe (AU) app within two weeks of its initial release in Australia. Similarly, downloads for the Indian tracing app Aarogya Setu have crossed the 114 million mark. Generally, the number of downloads for an app is used as an indicator for user acceptability. However, we argue that this is not a sufficient metric to gauge the impact or effectiveness of a contact-tracing app. We require additional information such as the number of close contacts identified through the data captured by an app and the associated false positive/false negative rates. Typically, health officials undertake manual contact tracing in conjunction with the app data to find out the close contacts for an identified case. Three distinct possibilities exist in this regard:
a) The contacts identified by the app matches exactly with the manual tracing process; App has achieved its objective, and the data is used as confirmation.
b) The contacts identified by the app are more than those identified by the manual tracing process; App has achieved its objective and proved its effectiveness.
c) The contacts identified by the app are less than those identified by the manual tracing process; The performance of the app is questionable.
However, as of this writing, there is no openly available data that can be used for impact analysis and measuring the effectiveness of any of these apps.


 Authors in \cite{T&P} has discussed several concerns associated with the use of these contact tracing apps. We discussed concerns related to user data privacy  and security in Sections  \ref{sec:Privacy} and \ref{sec:Attacks}. In this section, we focus on some of the additional concerns from user's perspective.

\subsection{Battery Usage}
Excessive battery consumption is a recurring problem for mobile apps. Mobile battery consumption is affected by many factors such as app processor utilisation, frequency, the scale of data management, and the number of messages exchanged, etc. Most tracing apps rely on the BLE communication protocol to exchange information with peers, while other apps use regular cellular or Wifi connections to communicate with servers. The BLE protocol allows the app to exchange a small amount of data with peers periodically. On the other hand, communication with the server relies on traditional secure application protocols, e.g., HTTPs. The main impact on the battery utilisation for these protocols is related to the number of information exchanges with the server. Table \ref{table:Battery} shows the comparison of different factors that affect mobile battery consumption. For the apps that rely on a centralised architecture, a fixed-sized message is retrieved from the server periodically to get a new TempID.

In contrast, for the decentralised version, there is no periodic data retrieval during the operation stage. Data is only downloaded once the server publishes the seeds uploaded by a positive case. The app nevertheless checks for any new download after every 24 hours. The upload rate of decentralised apps is lower than centralised counterparts. In decentralised cases, the upload consists of all seeds used during the past 21 days, whereas the centralised apps upload all encounter messages captured in the last 21 days. Centralised apps perform better in terms of processing at the devices as this involves minimal processing compared with the decentralised apps that need to generate and maintain seeds (after every 1 hour) and chirps (after every 1 min). For the hybrid architecture, devices generate EphIDs as well as two PETs for each received EphID. Hybrid apps upload the highest amount of data to the server: PETs from identified positive cases and Query PETs from all other users, collected for the purpose of checking their risk status. On the other hand, download from the server is lower compared with other architectures.

Battery consumption is also affected by the execution aspect, as an app that runs in the foreground demands more power than an app that runs in the background. Typically this design choice is dependant on the operating system support for these apps. As a case in point, CovidSafe and TraceTogether IoS apps face these documented issues when the app is running in the background~\cite{bluetraceWhitepaper}. Google and Apple, the two leading smartphone operating system providers, have teamed up to provide exposure notification APIs that improve the tracing application integration with the operating system. These APIs are  expected to improve the apps' development process and reduce power consumption.

\subsection{Compatibility of OS versions and different apps}
Developing an app that works in all smartphone models is not a trivial task. Smartphones run different versions of the operating system (OS), Android and iOS being the two dominant OSs. Most of the tracing apps have been developed for the newer OS versions; for example, both TraceTogether and CovidSafe require iOS version 10 or higher. TraceTogether requires Android 5.1 or higher, while CovidSafe works on Android 6.0 or higher. CovidSafe requires the newer OS versions for security reasons and improved Bluetooth capabilities \cite{covidsafeHelp}.

A secondary concern is cross-app compatibility. Consider a case when a user of an app released for a specific geographic region travels to areas where another app has been deployed. It is not clear how an app would behave if a second tracing app is installed on the same device due to the architectural differences discussed earlier. 

\subsection{Consent withdrawal}
\label{sec:consent}
Consent withdrawal refers to the ability of a user to stop participating in data sharing. Consent withdrawal provides a guarantee that users can delete their data or eliminate their existence from the tracing app ecosystem whenever they wish.
We consider next the consent withdrawal process at various stages of the app operation.
\begin{itemize}
    \item \emph{Data collection phase}: Since the encounter data is only locally stored at the user devices (for a limited period of 21 days for instance), deleting an app implies the removal of  all encounter data from the device immediately without transfer to a server. Additionally, for the centralised architecture, app deletion is assumed to result in the removal of all personal data captured at the registration stage. However, as the encounter data exchange is symmetric, other active users would have encounter messages/chirps stored in their devices for 21 days. Similarly, for the hybrid architecture, if a user withdraws consent and deletes the local data, PETs generated using their broadcasted EphIDs would remain in the local storage of contacts.

    \item \emph{Data already uploaded to the server}: If a user voluntarily uploads data to the server after testing positive, and subsequently wants to exercise consent withdrawal, can the system support it? In the centralised architecture, the uploaded data consists of encounter information from all close contacts of the positive case. The server processes this data and then alert close contacts. As this contact data is not transferred to the close contacts, the server can easily remove the uploaded data as it has already been utilised.  
    In terms of systems, the deletion of such data and provision of a receipt or feedback to the user is technically feasible.

    In the case of the decentralised architecture, the consent withdrawal process is more involved as the uploaded seeds from an infected user are made available to other app users for self-checking. If the infected user requests data deletion, the server can remove the stored data, but it is not clear how the seeds already transferred to other devices (and the reconstructed chirps) can be immediately expunged from these devices (transferred seeds are always removed after 21 days). As discussed earlier, the use of bloom filters for encoding hashed chirps at the server could provide an additional level of data protection (Section \ref{sec:Enum}). For the hybrid case, if a user withdraws consent after uploading the PETs, the server can remove the PETs as these are not transferred to other users.

    \item \emph{End of pandemic}: Most of the authorities managing tracing apps have indicated that the app data collected would be removed once the system gets de-activated at the end of the pandemic, often referred as the sunset clause, unless a user requests explicit earlier removal of their stored data. However, `end of the pandemic' is a vague term. Proposals such as DP-3T have an automatic mechanism for the removal of data from the server and the clients after a fixed period.
\end{itemize}

\subsection{Transparency}
There is a genuine public concern about the nature of information being collected from the user's smartphone and its usage by various parties. There are two essential approaches to achieving transparency. The first option is to make the source code of the app open. Publishing the app source code improves transparency and trust in the system, as the implemented privacy and security features can be scrutinised by the research and academic community. Although this is the preferred option, public source code is not a panacea to ensure security. Many risks arise from system configurations and the use of a wider system that is not readily observable from the source code analysis. It would also be useful to ensure that the code is subjected to periodic review and trusted third-party audits.

While transparency is a key to wider adoption by the end-users, it is essential to note that ultimately a degree of trust is required in the use of any mobile app.  This includes trusting the developers, the independent test and verification team, the operators and owners of the service, and importantly, the companies providing essential components such as the mobile phone operating systems.

Finally, all functional apps should be accompanied by a Privacy Impact Assessment (PIA). To the best of our knowledge, among the apps that have been covered in this article, only CovidSafe (AU) and DP-3T are accompanied by a PIA \cite{PIACovidSafe} \cite{lsts}.

\section{Future Directions}
\label{sec:future}
The current COVID-19 crisis has started the trend of using a virus tracking app at scale. We could have been better prepared for a pandemic of this proportions if these applications had been in place, fully tried and tested before the onset of the pandemic. There is already a flurry of research activities around the globe to develop the next-generation tracking applications ready for instant deployment should the world face a similar, or potentially even more dangerous pandemic. We consider some future research directions next.

In considering research areas for future tracing applications, it is probably best to consider near-term development, say the next five years, with more ``blue sky” type research pushed to post 5-year horizon. For the near-term, we identify the following as some important topics.
\begin{itemize}
    \item \textit{Improvement in proximity accuracy}: Current challenges related to proximity accuracy have been  highlighted in our earlier discussions. The Bluetooth technology was not designed with location or proximity determination as a critical part of the design process. However, the latest version of BLE released in 2020 has added features which will be helpful for next-generation apps. Additionally, new ``Bluetooth-like” protocols should be designed with an emphasis on location/proximity services during the inception phase. 
    Antenna design should be part of this process, allowing for not only distance accuracy but also direction-finding. Indeed, direction-finding is embedded as part of the Bluetooth 5.1 protocol albeit with low accuracy. Currently, no Covid19 tracing app  has included such direction capabilities within their proximity analysis. The use of time-of-arrival metrics could also be used as hardware technology improves, and sub-nanosecond timing becomes commonplace within devices. Indeed, several smartphone manufacturers have moved to embed a dedicated Ultra-Wideband (UWB) chipset  - specifically with Bluetooth proximity analyses in mind. Compared to the small bandwidth utilised by Bluetooth (2MHz), UWB utilises 500MHz, potentially resulting in cm-type precision for the proximity determination. 
     Research on fusing radio data with other location information garnered from other sensors embedded on the phone, such as location-designed WiFi, Enhanced-GPS, gyroscopes, accelerometers, and magnetometers is promising to improve the proximity accuracy. 
     These sensors could be aided by advances in software solutions such as advanced digital processing for location tracking and artificial intelligence-based algorithms. One can imagine new ``proximity” designs specifically aimed at integrating all relevant hardware, software, and protocol design into a single chip.
    \item \textit{A fully decentralised architecture for infection tracing}: One of the ``takeaway messages” from the COVID-19 crisis is that privacy concerns have to be addressed for  wider adoption by the public. None of the applications we have described above can be considered as encompassing a fully decentralised architecture - they all use a central server to differing degrees, usually under the control of a governing authority. Research on a fully decentralised system using some form of a peer-to-peer network to facilitate privacy-preserving information sharing amongst the user-devices should be pursued. 
    \item \textit{Artificial Intelligence-based algorithms}: AI algorithms are increasingly being used as the processing power within phones improves. The use of such algorithms in aiding the decision-making process of infection likelihood is obvious. Via metrics such as true infection identification, missed detections, and false-positive outcomes, the algorithms can become “live” in the sense that they dynamically adapt and self-improve in their reliability and accuracy. We anticipate much research in this area, especially with the integration of AI and the decentralised architecture.
\end{itemize}

In a post-5-year time scale, research outcomes are harder to predict. Much depends on the hardware development and the emergence of new technologies that are touted to become available. However, perhaps the most exciting of these emerging technologies lies in the quantum arena.
\begin{itemize}
    \item Quantum computing \cite{quantumComputing} is considered by many to be on the threshold of a breakthrough both in development and in commercialisation.  Future research on tracing solutions that exploit the use of the exponentially more powerful computing resources afforded by quantum computing should commence now. These could include quantum-based artificial learning algorithms and advanced Monte-Carlo or particle filter type tracking solutions.  In this paradigm, information from the phones will be sent to a central quantum computer for processing.
    \item Quantum sensing \cite{quantumSensing} is another area of current research that is also touted to bring major advances in the coming years. This technology exploits hypersensitivity embedded in quantum entanglement as a source of improved timing, network synchronisation, accelerometer accuracy, and location accuracy, to name a few. Clearly, all these issues could be brought to the contact tracing arena to maximise the overall performance and efficacy of the applications.
    \item Quantum communications \cite{quantumComm} is the most advanced of all the new quantum applications, at least in the sense of deployment. Commercial offerings in the quantum communications area already exist, and proof-of-principle deployment in space has already occurred. The major impact of this technology on tracing applications will likely be in the advanced communications security and enhanced privacy protections it offers. In principle, if properly deployed, security and privacy in future virus-tracking applications will be unconditional – hacking and unauthorised access to virus-tracking data will be made obsolete.
\end{itemize}

\section{Conclusions}
\label{sec:concl}
The COVID-19 pandemic continues to affect the way of life of everyone. The contact tracing apps are likely to play a vital role in aiding health authorities quickly identify individuals that may have been exposed to the virus. The imminent interest and adoption of tracing app technology will improve the tracing capability of health authorities; however, as this article highlighted, it is not a silver bullet. These apps still face many concerns from users, data protection agencies, and researchers. The main concerns are related to the user data management, potentially non-trivial false positive and negative instances, and the security and privacy issues of these apps. Guided by these concerns, this article presented an overview of the three common tracing app architectures: centralised, decentralised, and hybrid; and an overview of popular apps within these categories. Additionally, the paper focused on the privacy and security aspects, mapping attacks that could be possibly performed in each of the three architectures. This article also elucidates some other users' concerns regarding battery drain, compatibility, consent withdrawal, and transparency. Finally, we discussed some of the near and long term future research directions.

We note that each architecture has pros and cons, different attack models and protections, varying complexity of implementation, and operating costs. The adoption of a particular architecture, by a government, is based on familiarity with technology, integration with existing tracing processes, and ease of deployment. On the other hand, the adoption of an app by users is voluntary. The users have their due concerns regarding the privacy and security of their PII collected through these apps. The adoption rate by users can be increased significantly if complete transparency and legislative guarantees against misuse of data originating from this ecosystem can be assured by the authorities.

We also highlight that government agencies and service providers such as ISPs, and large corporations such as Apple/Google can already track people by using traditional apps and technologies such as  WiFi connections, the cellular communication tower areas, GPS navigation apps and a whole range of cameras deployed across cities. Since many users already install a myriad of apps on their phones or smart watches (games, social media, etc.) without knowing the security/privacy implications, installing a tracing app that primarily aims at helping in a noble cause of keeping the community safe from spreading the COVID-19 disease, in the opinion of the authors, should not cause undue concern.

We hope that this article will aid the research community to understand various technological and cybersecurity aspects of tracing apps and help users and agencies to make a more informed decision about the voluntary adoption of an app offered in their geographical areas.

\section*{Acknowledgements}
The authors thank Michelle Malaney and Xiaoyu Ai for their help in preparing this manuscript for publication. This work has been supported by the Cyber Security Research Centre Limited (CSCRC) whose activities are partially funded by the Australian Government's Cooperative Research Centres Programme.


\bibliographystyle{IEEEtran}
\bibliography{main}

\end{document}